\documentclass[a4paper,12pt]{article}
\usepackage{amsmath,amssymb,amscd}

\usepackage{graphicx}

\usepackage{hyperref}

\usepackage{subcaption}

\newcommand{\bea}{\begin{eqnarray}}
\newcommand{\eea}{\end{eqnarray}}
\newcommand{\beq}{\begin{equation}}
\newcommand{\eeq}{\end{equation}}

\newcommand{\tref}[1]{(\ref{#1})}

\newcommand{\tpre}[1]{}
\newcommand{\tprenote}[1]{}

\providecommand{\url}[1]{#1}
\providecommand{\href}[2]{#2}
\providecommand{\eprint}[1]{\href{http://arXiv.org/abs/#1}{\texttt{arXiv:#1}}}
\newcommand{\texpect}[1]{\langle #1 \rangle}

\newcommand{\cexpect}{\texpect{c}}

\newcommand{\kin}{k^{(\mathrm{in})}}
\newcommand{\kinexpect}{\texpect{\kin}}
\newcommand{\kout}{k^{(\mathrm{out})}}

\newcommand{\thalf}{\tau} %{T_{1/2}}
\newcommand{\Tmed}{T_\mathrm{med}}
\newcommand{\Ttotal}{T_\mathrm{tot}}
\newcommand{\Thalf}{T_{1/2}}
\newcommand{\Thalfdata}{T_{1/2,\mathrm{data}}}
\newcommand{\Thalfmodel}{T_{1/2,\mathrm{model}}}
\newcommand{\AS}{\thalf} %{\tau_\mathrm{AS}}

\begin{document}

%%%%%%%%%%%%%%%%%%%%%%%%%%%%%%%%%%%%%%%%%%%%%%%
%
% Comment this part out for submission to Scientometrics
%  and ALSO make sure you uncomment the \maketitle command below
%

\renewcommand{\thefootnote}{\fnsymbol{footnote}}

% Imperial/TP/14/TSE/2
\begin{center}
{\Large\textbf{Modelling Citation Networks}}
 \\[0.5cm]
 {\large \href{http://astro.qmul.ac.uk/directory/s.r.goldberg}{S.R.\ Goldberg}${}^1$, H.\ Anthony${}^2$ and \href{http://www.imperial.ac.uk/people/t.evans}{T.S.\ Evans}${}^2$}
 \\[0.5cm]
 (1) School of Physics and Astronomy, Queen Mary, University of London,
  \\ %327 Mile End Road,
  London, E1 4NS, U.K.
 \\[0.5cm]
 (2) \href{http://complexity.org.uk/}{Centre for Complexity Science},
Physics Dept., Imperial College London, \\ London, SW7 2AZ, U.K.
\end{center}

\providecommand{\title}[1]{}
\providecommand{\author}[1]{}
\providecommand{\institute}[1]{}
\providecommand{\keywords}[1]{\vspace*{12pt} {\centering Keywords: \parbox[t]{0.8\textwidth}{#1}}}
\providecommand{\subclass}[1]{} %{{\centering MSC: \parbox[t]{0.8\textwidth}{#1}}}
\providecommand{\PACS}[1]{} %{{\centering PACS: \parbox[t]{0.8\textwidth}{#1}}}
\providecommand{\date}[1]{}

%
% End of section to comment out for Scientometrics
%
% %%%%%%%%%%%%%%%%%%%%%%%%%%%%%%%%%%%%%%%%%%%%%%%%%%%%%%%%%%%%%%

\title{Modelling Citation Networks}
\author{S.R.~Goldberg \and H.~Anthony \and T.S.~Evans}
\institute{S.R.~Goldberg
\at School of Physics and Astronomy, Queen Mary, University of London,
  %327 Mile End Road,
  London, E1 4NS, U.K.\\ Tel.: +44-20-7882-5515
\and
H.~Anthony \and T.S.~Evans
\at \href{http://complexity.org.uk/}{Centre for Complexity Science},
Physics Dept., Imperial College London, SW7 2AZ, U.K.
}
\date{12th August 2014}

\begin{abstract}

The distribution of the number of academic publications as a function of citation count for a given year is remarkably similar from year to year. We measure this similarity as a width of the distribution and find it to be approximately constant from year to year. We show that simple citation models fail to capture this behaviour.  We then provide a simple three parameter citation network model using a mixture of local and global search processes which can reproduce the correct distribution over time.  We use the citation network of papers from the hep-th section of arXiv to test our model.   For this data, around 20\% of citations use global information to reference recently published papers, while the remaining 80\% are found using local searches.  We note that this is consistent with other studies though our motivation is very different from previous work. Finally, we also find that the fluctuations in the size of an academic publication's bibliography is important for the model. This is not addressed in most models and needs further work.


\subclass{91D30} %MSC codes

 %PACS codes

\end{abstract}

% %%%%%%%%%%%%%%%%%%%%%%%%%%%%%%%%%%%%%%%%%%%%%%%
%
% Don't forget this for journal, comment out for plain style
%
%\maketitle

% LIKEWISE must comment these footnote lines out for journal but leave in for plain style
\renewcommand{\thefootnote}{\arabic{footnote}}
\setcounter{footnote}{0}
%\newpage

%
% %%%%%%%%%%%%%%%%%%%%%%%%%%%%%%%%%%%%%%%%%%%%%%%

%MSC: http://www.ams.org/msc
%91D30  	Social networks

% PACS: http://www.aip.org/pacs
%05.40.Fb   Random walks and Levy flights
%05.50.+q   Lattice theory and statistics (Ising, Potts, etc.)
%75.10.Hk   Classical spin models
%89.75.-k Complex systems (for complex chemical systems, see 82.40.Qt; for biological complexity, see 87.18.-h)
%89.75.Da	Systems obeying scaling laws
%89.75.Fb Structures and organization in complex systems
%89.75.Hc Networks and genealogical trees
%89.75.Kd	Patterns
%89.65.-s	Social and economic systems
%89.65.Cd	Demographic studies
%89.65.Ef	Social organizations; anthropology
%89.65.Gh	Economics; econophysics, financial markets, business and management (for economic issues regarding production and use of renewable energy, see 88.05.Lg)
%89.65.Lm	Urban planning and construction

% ********************************************************
\section{Introduction}

A citation network is defined using a set of documents as vertices with directed edges representing the citations from one document to another document.  Examples include networks from patents (for example \cite{SBS08,clough2013transitive,CE14}) and court decisions (see \cite{FJ08,clough2013transitive,CE14} for an example)  but in this paper we will work with data from academic papers \cite{P65,P76} and our language will reflect this context.  Since citation networks capture information about the flow of innovations, understanding the large scale patterns which emerge in the data is of great importance.

The citation distribution, namely the number of papers with a given number of citations, is one of the simplest features and it has long been known to be a fat-tailed distribution \cite{P65,P76}, a few papers garner most of the citations. However the focus of our work is the more recent observation that the shapes of these distributions are surprisingly stable over time \cite{RFC08,EHK12, Goldberg}.  This is a feature that many simple models for the process of citation fail to capture, refer to model A.  Our aim is to produce a simple model which will enable us to understand the origin of this feature. In order to create a model close to the real world we will work with a real citation network derived from papers posted between 1992 and 2002 on the hep-th section of arXiv.  We will use this data to provide both key input parameters for our models and as a test of the output from our models.

In section \ref{sec:A} we analyse the key features of our real citation network.  Then, we define our models A to C in sections \ref{sec:MA} to \ref{sec:MC}. In each case we identify strengths and weaknesses, using the latter to provide motivation for improvements to the models.  Finally, we conclude and discuss further work in section \ref{conc}.

% ******************************************************************************
\section{Analysis of hep-th arXiv Data}\label{sec:A}

In this section we will describe the data and analyse its key features. In doing so, we will define our notation and analysis methods. The data comes from the hep-th section (high energy physics theory) of the arXiv online research paper repository \cite{cup} citation network from 1992-2002.  As this section only started in 1991, we suggest that the early data may have some initialisation effects. For example, the number of publications released per year increases rapidly at first, but after 1994 the rise is a more gentle one, see figure \ref{fig:data_papers_per_year}. Anecdotally, hep-th rapidly became the defacto standard for the field (it has always been free to both authors and readers with easy electronic access) and we suggest that essentially every paper in this field produced after 1994 is in our data set. This completeness gives us over 27,000 publications in our data, so we have enough information to draw useful statistical inferences. Different fields have a different number of papers published per year and different distributions for the number of references associated with these papers \cite{RFC08}. By creating a citation network model associated with one field alone (hep-th) we eliminate any bias in our model due to field dependence. Overall, while all citation data sets miss citations to documents outside the data set, we are confident that the post 1994 parts form a reliable single field citation network. Another by-product is that our data is open source allowing independent verification; we took our copy from \cite{cup}.

An important feature of arXiv is that papers are given a unique identifier when they are first submitted.  This identifier records the order in which papers are submitted within a particular section, a larger number indicates a later submission, and the year and month of this first submission are also simply encoded in the identifier. For instance \eprint{hep-th/9803184} was submitted in March 1998 just after \eprint{hep-th/9803183} but just before \eprint{hep-th/9803185}. It is this first submission date which we take to define the publication date of each article.

% ----------------------------------------------------
\subsection{Number of Papers Published Per Year} \label{NPPPY}

\begin{figure}[htb]
\centering
\includegraphics[width=0.7\textwidth]{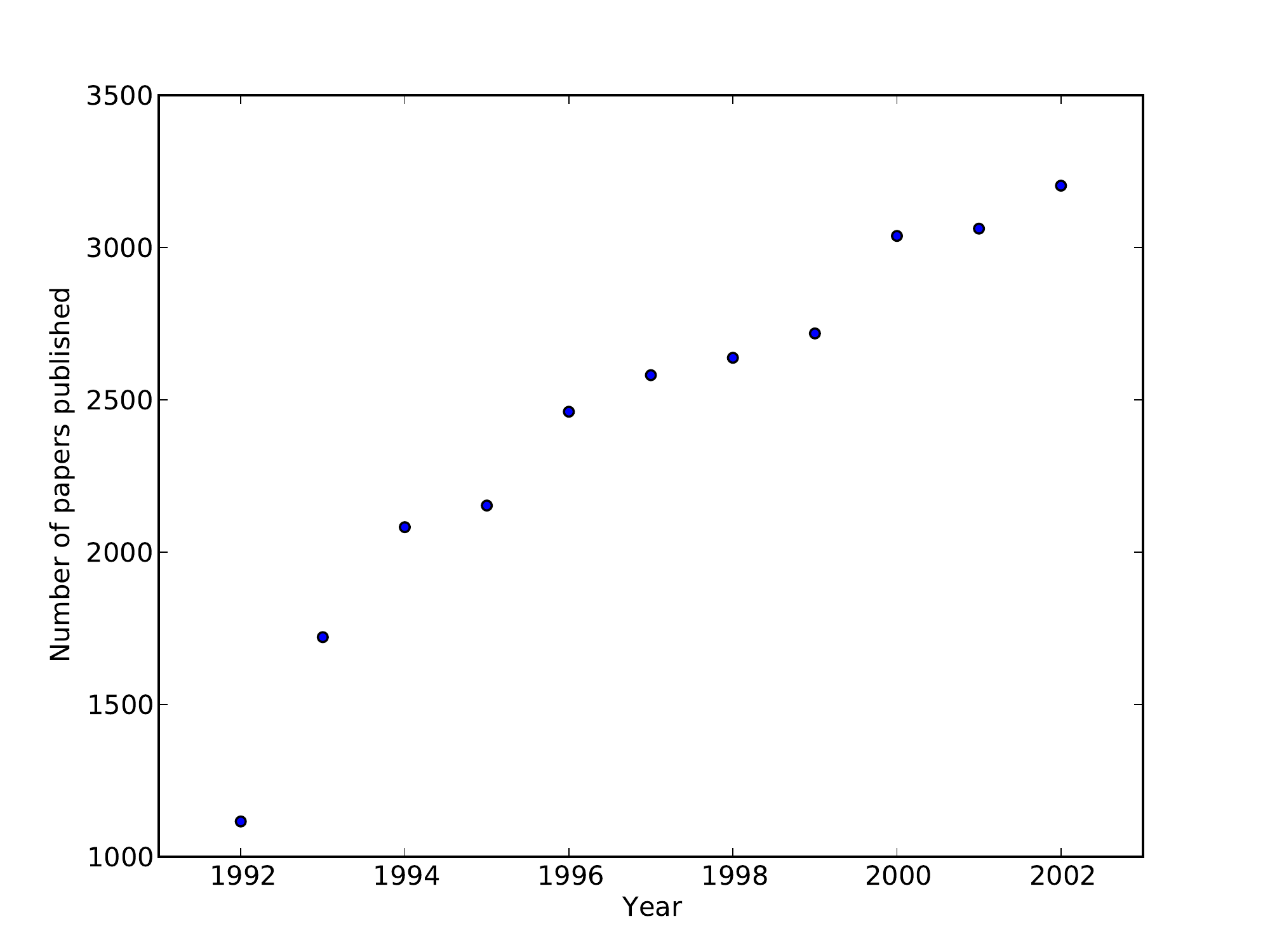}
\caption{This is a graph of the number of papers published in a given year against year for the hep-th arXiv dataset. An increase is clear, the number of papers published per year increases in an approximately linear fashion, it is not constant. The number of publications released from 1992 to 2002 approximately triples from 1116 to 3203 papers. When year to year comparisons are made from the hep-th data to our citation network models, it is important to incorporate literature growth with year. }
\label{fig:data_papers_per_year}
\end{figure}

The increase in the number of publications released per year is shown in figure \ref{fig:data_papers_per_year}. This is expected intuitively and in the literature. Intuitively, because we live in a more highly connected era where an increasing number of publications are written. Many authors, e.g.\ \cite{AMTC}, find that the number of papers published  increases with time. It increases approximately linearly after the initial start of the arXiv, figure \ref{fig:data_papers_per_year}. Therefore, when making year-to-year comparisons from the models to the hep-th data we compare the same number of publications in both cases to avoid any bias. For example, if we compared some quantity from our model in year 10 to the corresponding $10^{\text{th}}$ year in the hep-th dataset, 2002, there may exist a bias due the hep-th citation network having up to 3 times more papers/nodes in a year. We treat this growth in paper numbers as a separate external aspect of the citation process and do not attempt to model this growth.

By analysing citation data from the arXiv between 1992-2002 we are observing the network close to when it started\footnote{We omit year 2003 from our analysis because it is incomplete.}. There may exist some initialisation effects that requires a settling down period. For example, from 1992 to 1994 the number of publications released per year increases rapidly, then approximately linearly after 1994, figure \ref{fig:data_papers_per_year}. This initial increase is explained by two effects: the effect of the arXiv becoming a more popular network to reference and the quantity of literature published per year increasing generally, irrespective of the citation database, due to increase in the spread of knowledge. After the arXiv had gained popularity, after 1994, only the second effect controls the increase.

In our models we create 28,000 nodes, the same number of nodes as the hep-th arXiv citation network (27,000 nodes) plus 1,000 nodes. Then, we deleted the first 1,000 nodes and directed edges to those nodes. This corresponds to deleting the first year (as 1,000 papers corresponds to approximately one year, figure \ref{fig:data_papers_per_year}). This replicates the hep-th data.

% ----------------------------------------------------
\subsection{Citation distribution}

The citation or in-degree distribution is one of the most fundamental features of any citation network and is the focus of our attention here. It is well known that that these are invariably fat-tailed distributions with the tails often described in terms of power-laws.  Such distributions are for whole networks formed from all the citations between papers in a single data set, so papers published in many different years and different subjects.

Our principle concern is in the shape of these distributions for subsets of papers published in the same field and in the same year which show a similar fat-tailed distribution.  We are not concerned here about the precise description used for the shape of such distributions. All we require is a functional form of few parameters which we can use to fit these fat-tails, one that has proven effective on real data elsewhere and which we can test on our hep-th arXiv data. In this way we can capture the behaviour of these fat-tails from papers published in the same field and the same year in just a few parameters.  Our aim is to study the time evolution of these tails through the time evolution of our chosen few parameters. From this we can address our key question: what are the basic features required in a citation model in order to reproduce the correct time evolution of the citation distribution?

Our approach is inspired by the results of \cite{RFC08} when fitting their citation distributions for papers in the Web of Science data.  Their results suggest that using a lognormal distribution is a good way to capture these citation tails. In particular, we infer from their results that the tail of the citation distribution for papers published in one field and in a single year have a simple time evolution which is easily described in terms of a lognormal distribution. We note that the lognormal form has been effective in another study of citation data \cite{EHK12, Goldberg} and has been shown to be one of the better fitting forms in many other types of data with fat-tailed distributions \cite{ISI:A1990EV52100026,ISI:000088918500004,M04,ISI:000271983500002,Include1,N10}. Debate \cite{WER12} on other aspects of the work of \cite{RFC08} does not affect our approach here.

We use the approach discussed in \cite{EHK12, Goldberg}. First we select all the papers published in one period of time, typically one calendar year.  We then calculate the citation count for this year, the in-degree for the corresponding vertex in the citation network.  From this we get the average citation count for this subset of papers and we denote this as $\texpect{c}$.  Finally we fit the tail of the citation distribution, that is for $c>0.1\texpect{c}$ the statistical model for the number of papers with $c$ citations is
\begin{eqnarray}
 n(c)
 &=&
 (1+A) N \int_{c-0.5}^{c+0.5}  dc \; \frac{1}{\sqrt{2\pi} \sigma c }
 \exp\left\{ -\frac{(\ln(c/\texpect{c})+(\sigma^2/2)-B)^2}{2\sigma^2 } \right\}
 \, .
 \label{lognormal}
\end{eqnarray}
Here $N$, the number of papers in our time period, and $\texpect{c}$ are fixed by the data.  This leaves us with three parameters to fit: $A$, $B$ and $\sigma$. If we had a perfect lognormal then $A=B=0$.  As noted in \cite{EHK12}, if the number of zero and low cited papers, those with $c \leq 0.1 \cexpect$, follows a different distribution, then neither $A$ nor $B$ will be zero. This is typically found to be the case but this is not the focus of our work.  The only output parameter we use is $\sigma$. This is a measure of the width of the $\ln(c)$ distribution, it is not the square root of any variance measure of the citation distribution itself. Thus by working with the normalised citation count $c/\texpect{c}$ we account for one of the major differences between papers published at different times, namely that older papers have more citations. What we are focusing on is on the temporal evolution of the width of the distribution. Since we have a fat-tailed distribution, it makes sense to work in terms of the width of a $\ln(c)$ distribution.  Note $\sigma^2$ is not simply the statistical variance of the $\ln(c)$ distribution because we only use a sample of the distribution\footnote{We only fit the lognormal to the in-degree distribution of the hep-th arXiv for reasonably well cited papers with $c>0.1 \cexpect$ to estimate $\sigma$.}.

%The arXiv hep-th field as a constant $\sigma^2 =1.78\pm 0.14$ has a higher than average $\sigma^2 =1.3$ found by Radicchi et al.\ \cite{RFC08} and others \cite{EHK12, Goldberg}. This could be because hep-th (theoretical high energy physics) is a field covering both theoretical and experimental physics, therefore there is a large spread in the in-degree distribution. Nevertheless if our model Can replicate the $\sigma^2$ plot of this dataset, one of the highest $\sigma^2$ plots observed, we expect it to be good of others.
%

\begin{figure}[h] %[H] %or [h] %[H]
\centering
\includegraphics[trim = 0cm 0.5cm 0cm 1.5cm, clip = true, scale=0.5]{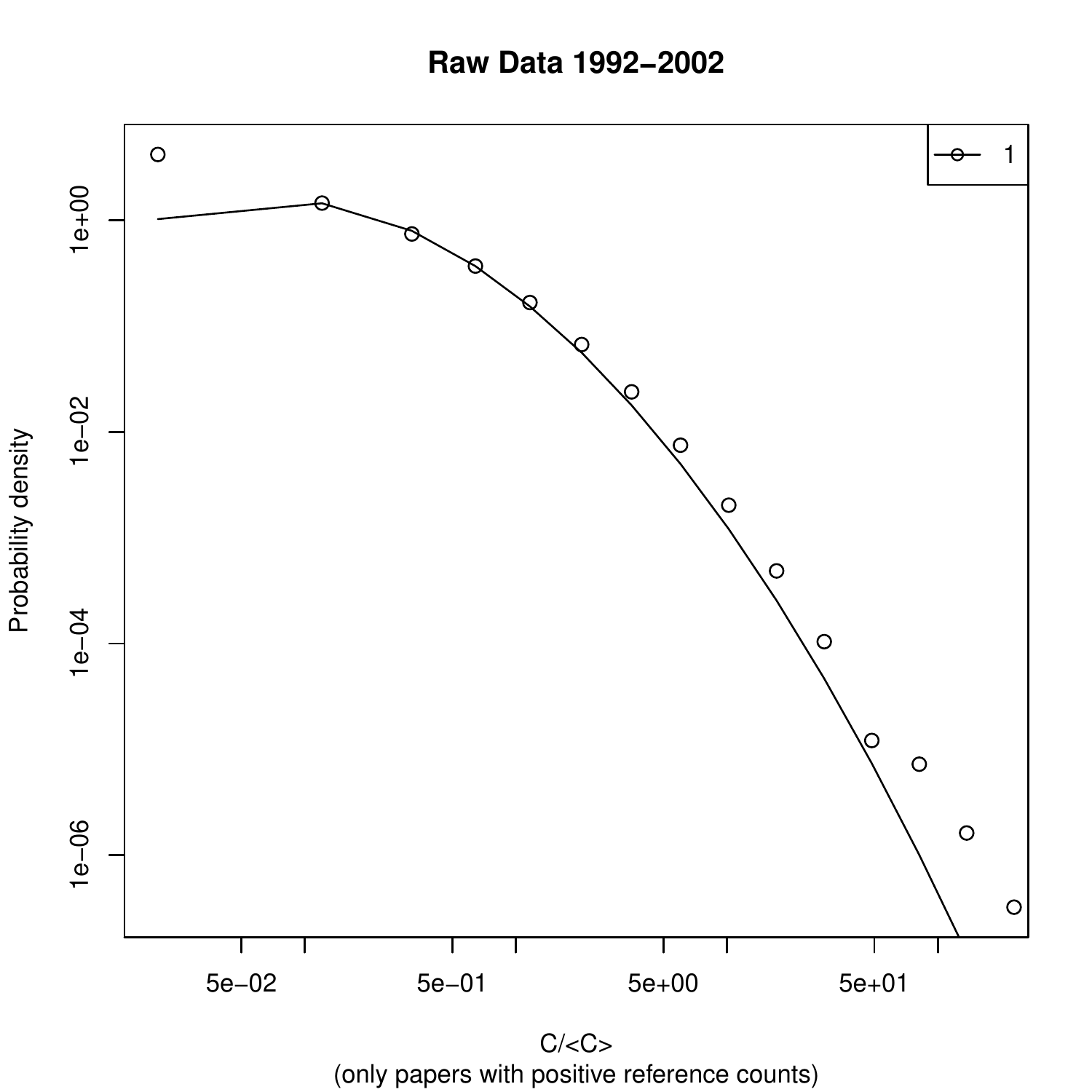}
\caption{The citation distribution for all papers in the hep-th arXiv repository between 1992 and 2002. The points are the counts from logarithmic sized bins and plotted in terms of $c/\cexpect$ where $\cexpect$ is the average citation count taken over all papers. The line shows the form of the best fitting lognormal curve, equation \tref{lognormal} \cite{Goldberg,EHK12}. The width of the distribution is $\sigma^2 = 1.78\pm 0.14$, large compared to the $\sigma^2 \approx 1.3$ found by \cite{RFC08, WER12}. }
 \label{dataindeg}
\end{figure}

We first look at the fat-tail of the distribution for all the papers in our data set, shown in  figure \ref{dataindeg}. We used logarithmic binning excluding zero cited papers\footnote{The bin scale was chosen to ensure there were no empty bins below the bin containing the highest citation values.}. We used the width of the in-degree distributions of papers published in the \textit{same year} for  years 1992-2002 as a measure to compare our model to the hep-th network. We find the width $\sigma^2 = 1.78\pm 0.14$ of the entire hep-th arXiv data set large compared with the literature where $\sigma^2 \approx 1.3$ \cite{WER12}.

\begin{figure}[h] %[H] %or [h] %[H]
\centering
\includegraphics[scale=0.4]{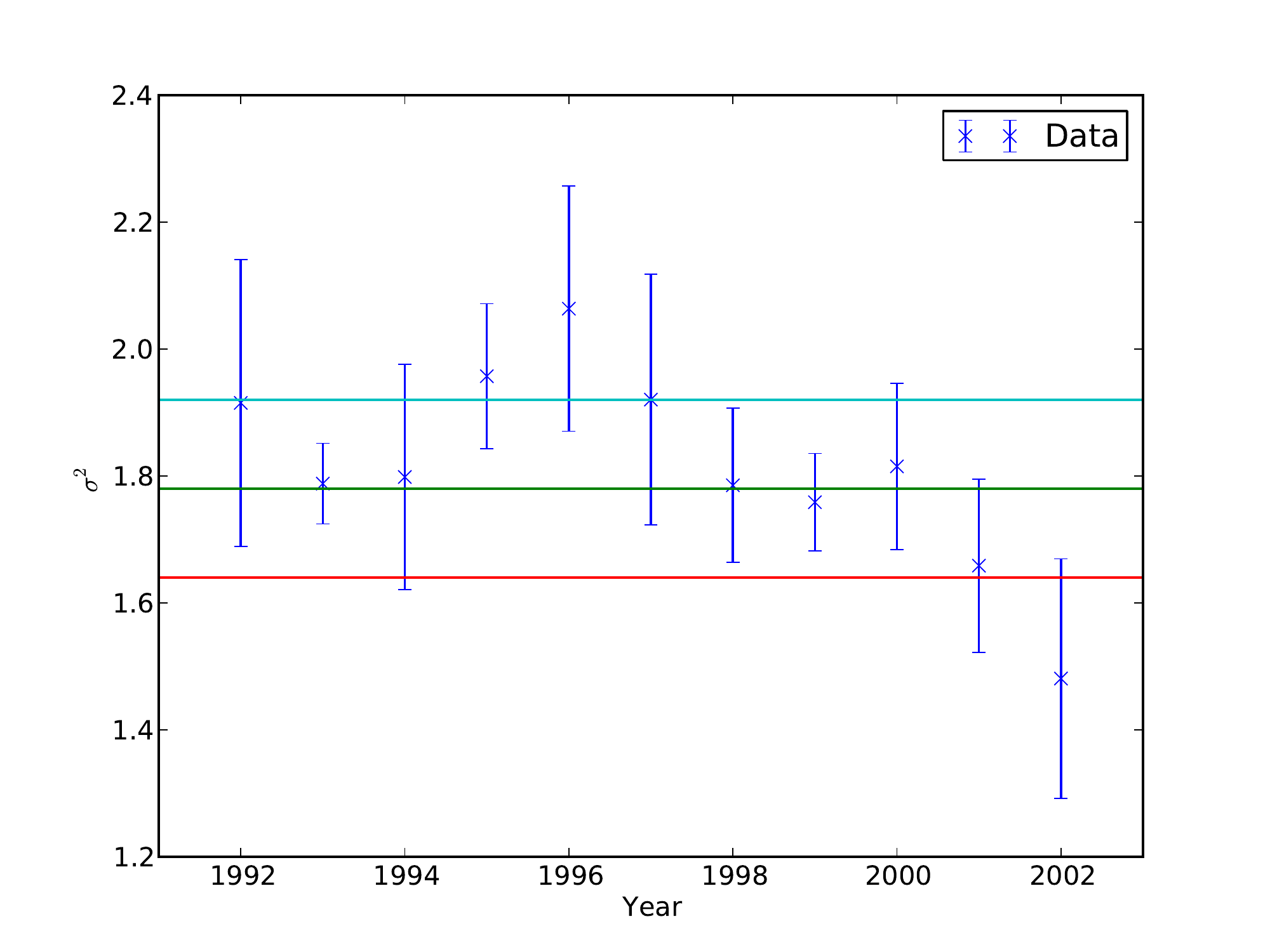}
%\begin{figure}[h] %[H]
 %       \centering
\caption{The $\sigma^2$ plot is a plot of $\sigma^2$ against year for the hep-th data. To calculate this we plot 11 in-degree distributions (corresponding to citations gained by papers published in years 1992-2002), each against the normalised citation count. We fit a lognormal to each of these distributions. Each lognormal gives a signature width of the distribution, $\sigma^2$. We verify the Evans et al.\ conjecture that $\sigma^2$ is constant with time. We find all $\sigma^2$ and their error bars lie within the $\sigma^2$ $=1.78\pm 0.14$ of the entire network, horizontal lines. The last two years are lower than the rest because they have three times as many publications as years 1992 or 1993, they pull the average down. 2002 has $\sigma^2$ much lower than other years and the trend from 2001 is downward. This is because the dataset used half of 2003 (we analysed up to 2002 and omitted the last year because it was incomplete) therefore the last year or so had very little time to gain citations, less than a year. Hence, most papers had few citations and the spread, $\sigma^2$, was smaller.}
\label{datas2}\label{datasto}\label{datas2zero}
\end{figure}

If we now we fit our lognormals to the in-degree distribution for papers deposited on arXiv in same calendar year, we find a reasonable fit. Moreover, we confirm the results of \cite{EHK12} that these fits are approximately the same with our measure of the width of the distribution, $\sigma^2$, remaining roughly constant. In figure \ref{datasto} most of the error bars lie within one standard deviation of the $\sigma^2 =1.78\pm 0.14$ for the \textit{whole} data set, we use this as an `average' of the distribution in figure \ref{datasto}. At the very least there is no evidence for any systematic change in the width over time. The challenge now is to find a simple model which can reproduce this effect.

% --------------------------------------------------------------
\subsection{Zero Cited Papers}\label{ZCP}

The primary focus of our work is to ensure that we get the correct time evolution for the width parameter $\sigma$ for the tail of the citation distribution. However there are large numbers of low cited papers not included in that aspect of the analysis.  To make sure our models give a realistic result for the low cited papers we simply look at the proportion of zero cited papers in the citation network, $z$, which can be up to 40\% of the data set \cite{Goldberg,WER12}.

% --------------------------------------------------------------
\subsection{Length of Bibliographies}\label{out-degree}

\begin{figure}[h] %[H]
\centering
\includegraphics[width=0.6\linewidth]{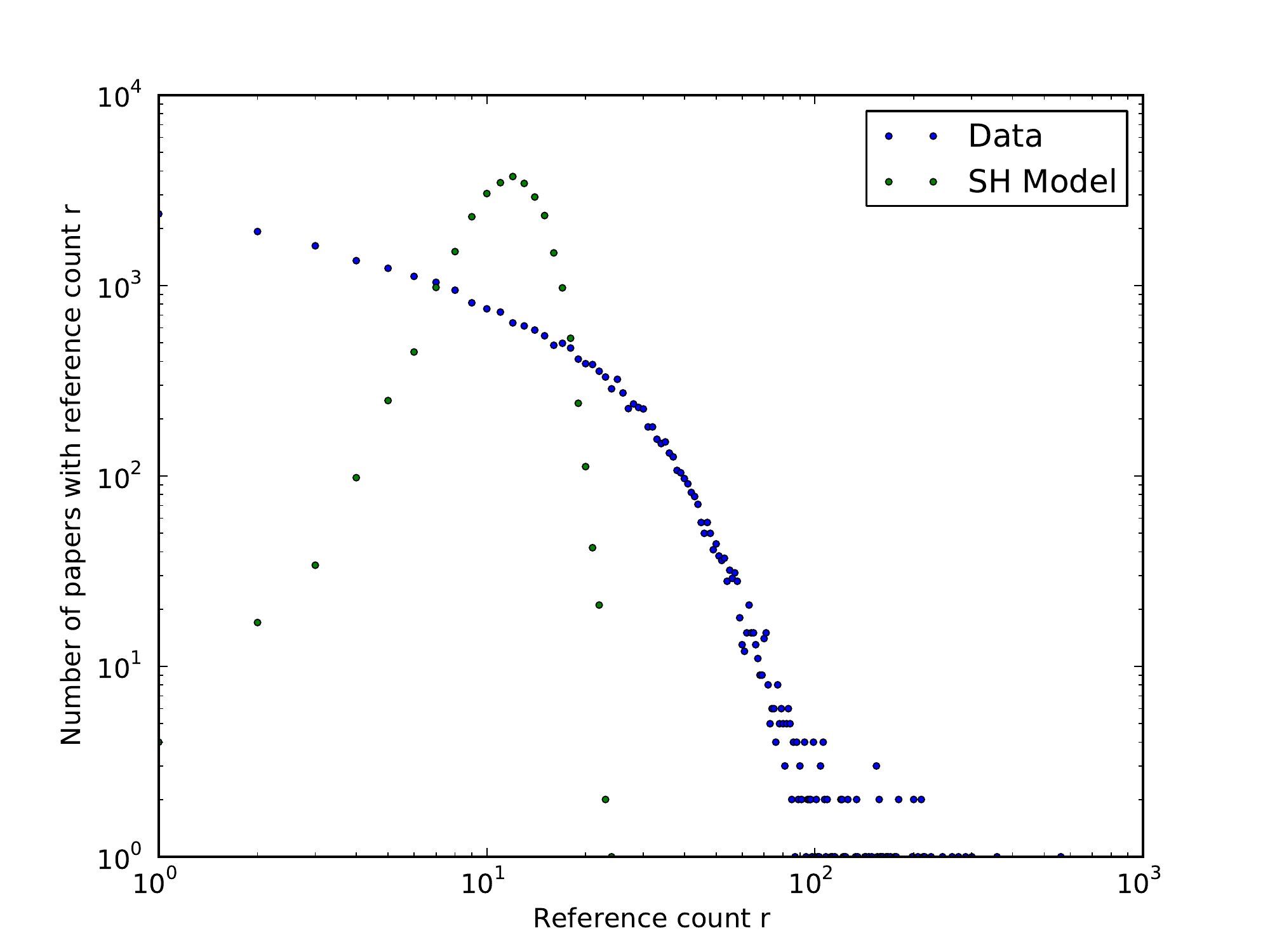}
\caption{This is a plot of the out-degree distribution, number of papers with reference count $r$ against reference count $r$, on a log-log plot for the data, in blue. Superimposed in green is a plot of the out-degree distribution used in our models, a normal distribution with the same mean 12.0 and standard deviation 3.0 as the whole hep-th data for the same number of publications, 27,000. This normal is clearly not a good fit for the out-degree distribution.}
\label{fig:model_and_data_outdeg}
\end{figure}

While the citation count or in-degree of a paper is an important measure, for citation network models we also need to understand the length of the bibliographies, the out-degree of papers. The out-degree distribution is again a fat-tailed distribution though not nearly as broad as the in-degree one, as figure \ref{fig:model_and_data_outdeg} shows and as noted in the literature \cite{citeulike:1007417}. We find that a lognormal distribution gives a reasonable description of the out-degree data, as seen in figure \ref{fig:model_and_data_outdeg_lognorm}.

Many models choose the simplest approach and use a constant out-degree \cite{AMTC,ISI:000187183600031}. Another approach would be to duplicate the actual out-degree distribution as in \cite{ISI:000302837300019}.  As we show in appendix \ref{logout}, changing the distribution of the length of bibliographies (number of references made by a paper) alters the results of the final model, see figure \ref{fig:model_and_data_outdeg}, so this is an important point. Refer to \cite{GoldC4} for more discussion. In particular we found that our model only gave good fits for the citation distribution when we used a normal distribution for the out-degree distribution, with a mean 12.0 and standard deviation 3.0 equal to that found from the actual data.

%, derived from the hep-th data for the reasons described in \ref{subsubsec:introOutdeg}. Here we observe the out-degree of hep-th, section \ref{O1}, and the shortcomings of current citation network models, section \ref{O2}.
%
%In our model the number of references created by each new node fits a normal distribution. How does this compare to the data? To investigate this we plot the number of papers with reference count $r$ against reference count $r$ (the out-degree distribution) for the data and the model (normal distribution with mean 12.0 standard deviation 3.0). Running this distribution 27,000 times is equivalent to creating the references of 27,000 nodes, figure \ref{fig:model_and_data_outdeg}.

We now turn to look at how three different models perform.

% ******************************************************************************
\section{Model A: The Price model}\label{sec:MA}\label{ModelAmot}

An obvious place to start is with the Price model of cumulative advantage \cite{P65,P76}. In the Price model new documents cite existing publications in proportion to their current number of citations, cumulative advantage, and this is well known to generate a fat-tailed distribution for citations. In our version, model A, at each step one new publication is added to the $N$ publications in the network.  The number of papers in its bibliography, $r$, is chosen from a normal distribution, described above. The new publication then makes references to different pre-existing publications in one of two ways: with probability $p$ it uses cumulative advantage (referencing a publication chosen in proportion to its current citation count) otherwise with probability $(1-p)$ it chooses uniformly at random from all existing documents.  References for the new paper are chosen repeatedly until $r$ distinct existing papers have been found and these then define the $r$ new links in the citation network.
The probability of making a link to an existing node $i$ is $\Pi_A(i)$ where\footnote{There is a small correction to this form due to the possibility of that the same vertex $i$ could be chosen more than once which is excluded in the actual model.}
\begin{equation}\label{AttachProbA}
\Pi_A(i)= \frac{p \kin_i}{\kinexpect N} + \frac{(1-p)}{N} \, .
\end{equation}
Here $\kin_i$ is the in-degree of node $i$, $\kinexpect$ the average of the in-degree and $N$ is the total number of papers.

%\tnote{The first term on the right hand side of the equality \ref{AttachProbA} refers to the probability of cumulative advantage, the second term refers to the probability of uniform attachment. Figure \ref{Amodel} is a visualisation. A mean-field approximation turns out to be exactly solvable in the infinite time limit \cite{KR01,N10} and in turn this is a very good approximation to numerical simulations of the model.}

%%\subsection{Motivation for model A}\label{ModelAmot}
%%There are two reasons why we set up model A as described above. Firstly, a citation network model must replicate the long-tail signature in-degree distribution of real citation networks, e.g. figure \ref{dataindeg}. These approximately follow power-laws or lognormals, section \ref{measures}. Why does the in-degree distribution have a long-tail?
%
%\subsection{Results}
%
%\subsubsection{Method of Determining and Value of \p}

\begin{figure}[h] %[H]
\centering
\includegraphics[scale=0.4]{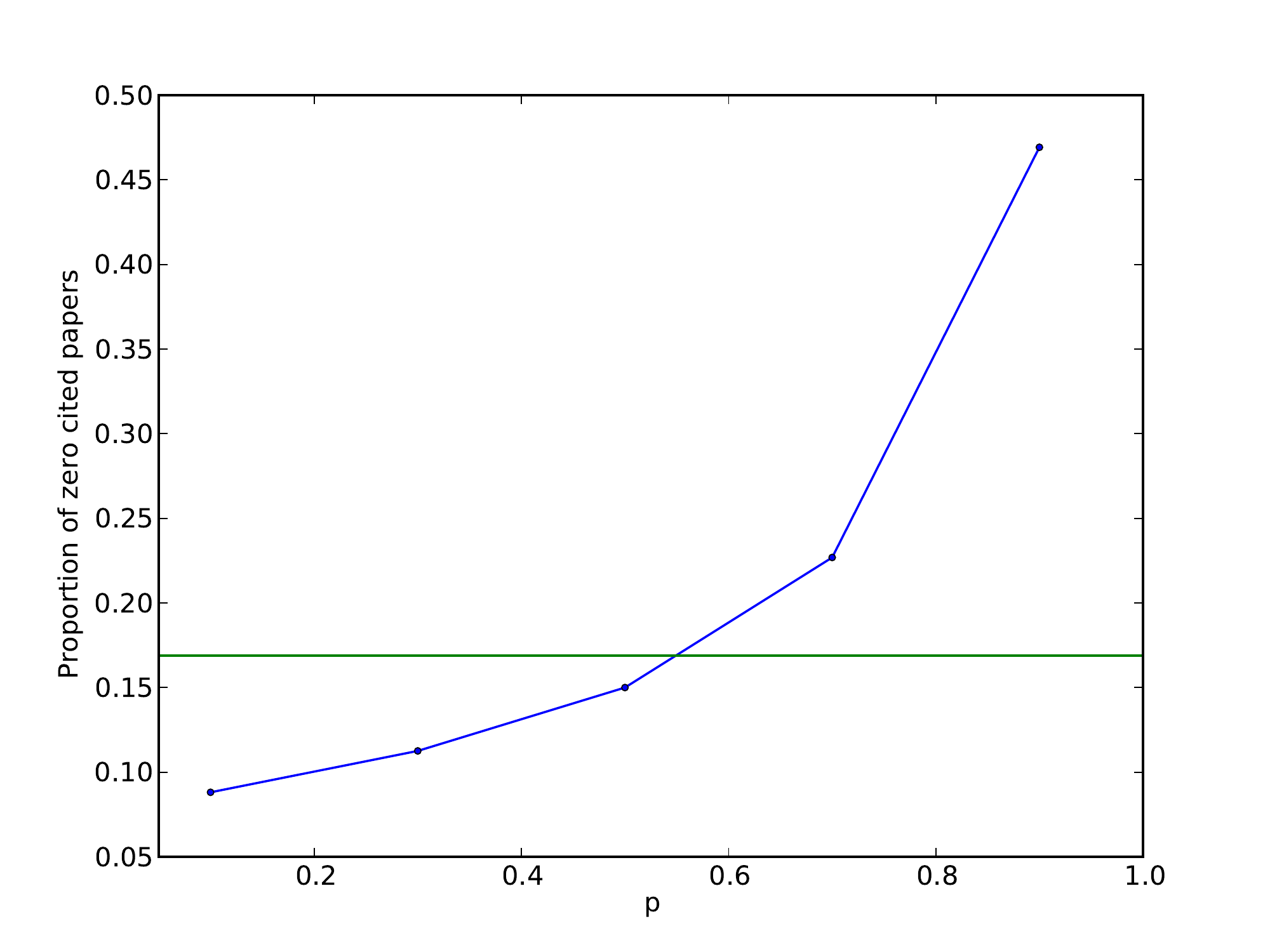}
\caption{The blue line is a plot of the proportion of zero cited papers, $z$, in model A for a given probability of cumulative advantage $p$, against varying $p$. As expected, as $p$ decreases uniform random attaching increases, this gives more opportunity for zero cited papers to attach. The green horizontal line is a constant line where the $z$ is 0.169, the proportion of zero cited papers in the hep-th arXiv data set. Where the lines cross is where the proportion of zero cited papers in model A is equal to that of the data. This occurs at the desired $p$, 0.55.}
\label{Avaryp}
\end{figure}

To determine the parameter $p$ we use the observation that the proportion of zero cited papers over all years in hep-th is $z=0.169$ as a constraint on our model. We vary $p$ in our model A and find the overall $z$ in the model as shown in figure \ref{Avaryp}. We then chose the parameter $p=0.55$ so that model A gives the same  number of zero-cited papers overall as found in our data\footnote{The full Price model may be solved exactly within the mean-field approximation in the infinite time limit.  Those solutions are very close to numerical results found for finite sized simulations such as ours.  For $\texpect{\kin}=12.0$, these formulae give $z=0.156$ if $p=0.55$ and we find we need $p=0.59$ in order to get the same value of $z=0.169$ found in our data.  However, we remove the first thousand papers created in our simulation so we do not expect an exact match with the theoretical expressions.}.

\begin{figure}[h] %[H]
\centering
\includegraphics[trim = 0cm 0.5cm 0cm 1.5cm, clip = true, scale=0.5]{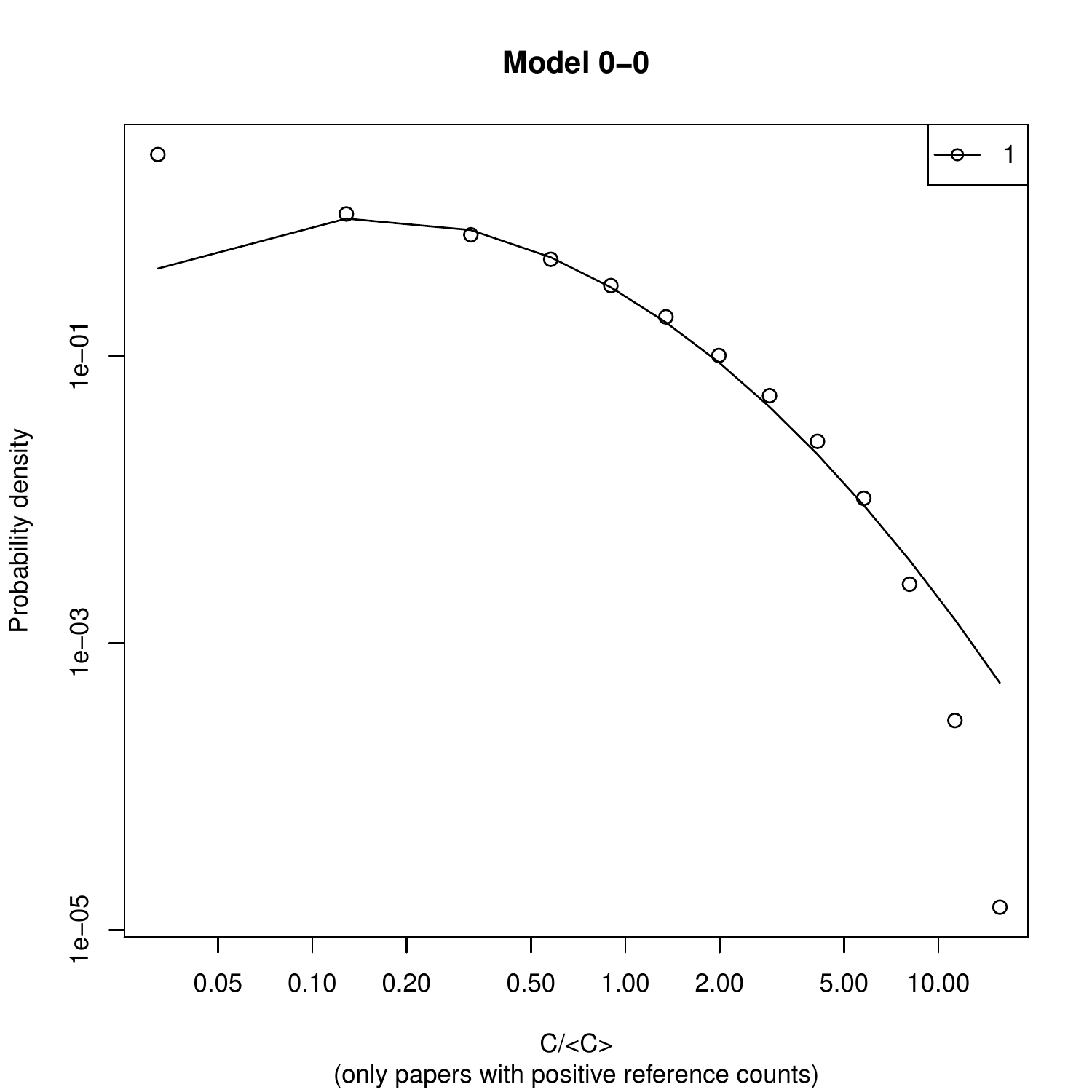}
\caption{This is the usual in-degree distribution (for normalised citation counts) for all years with a lognormal fit for model A. The axes are log-log, probability density against normalised citation count. We observe the long-tailed in-degree distribution goes up to approximately 20 (120 citations). The corresponding plots for the hep-th data goes up to 50 (500 citations), figure \ref{dataindeg}. This implies the width is not large enough. Another factor needs to increase the $\sigma^2$ values, next section.}
\label{indegma} \label{indegmaun} \label{ModelAunall}
\end{figure}

Turning to the long tail of the citation distribution, we first analyse the in-degree distribution for \textit{all} years for which model A and cumulative advantage are designed to give the expected long-tail. This we find but the overall width of the tail, figure \ref{indegma}, is $\sigma^2=1.05 \pm 0.18$. This significantly different from the data's $\sigma^2=1.78 \pm 0.14$, see \ref{ModelAunall}.

%%%

\begin{figure}[h] %[H]
\centering
\includegraphics[scale=0.4]{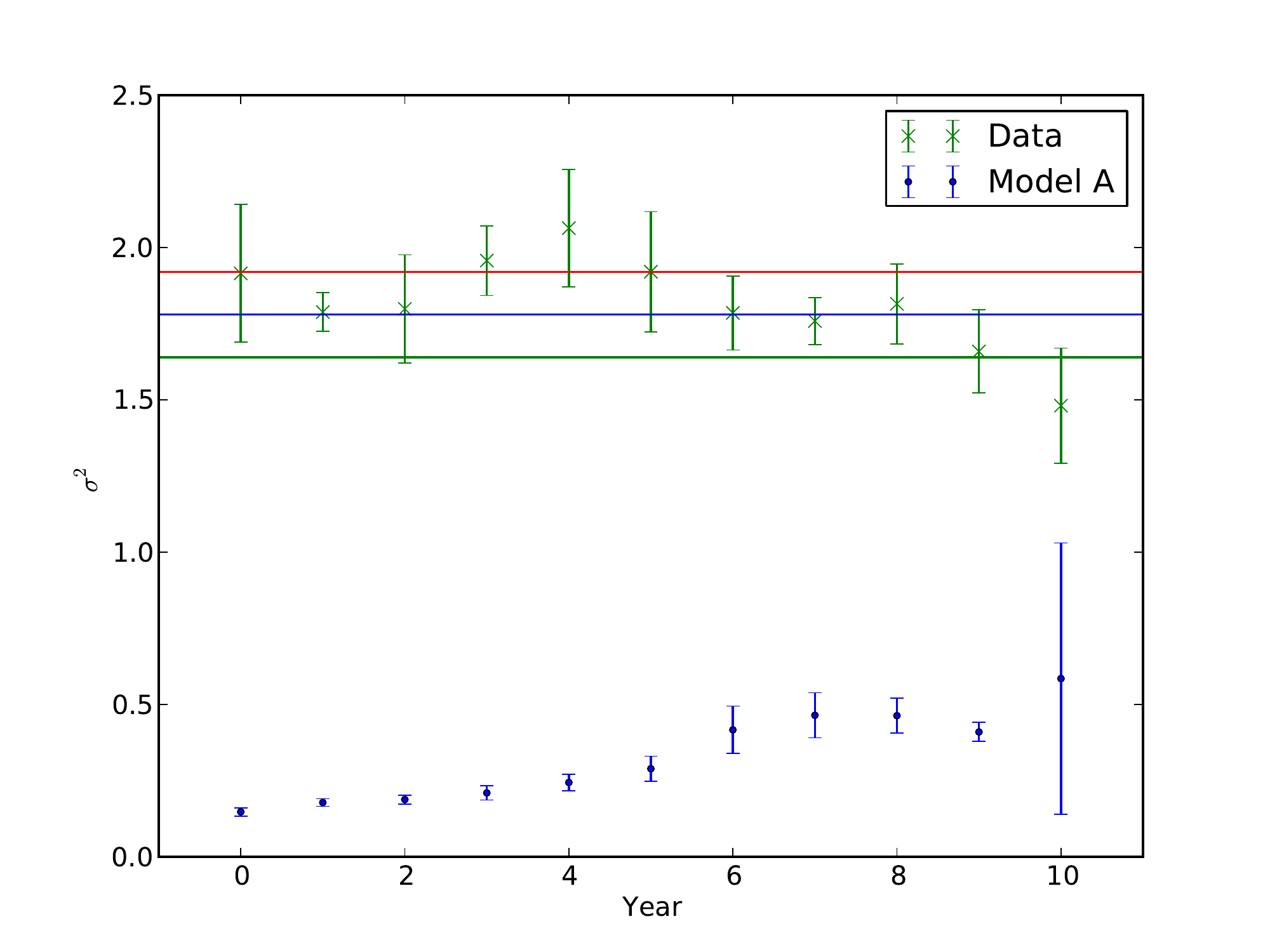}
\caption{This is the $\sigma^2$ plot, the plot of $\sigma^2$ against year for both the data, in green, and model A, in blue. $\sigma^2$ is the width associated with the lognormal fitted to the in-degree distribution for papers published in the same year. On this plot the data's years from 1992 to 2002 are relabelled years 0 to 10 respectively. The data's average and one standard deviation are plotted as horizontal lines $1.78\pm 0.14$. Model A and hep-th's $\sigma^2$ values are very different. }\label{ModelA}\label{sigma2dvma}\label{1992s2ma}
\end{figure}

We are more interested in the shape of the in-degree distribution for papers published in the same year.  Using $p=0.55$ we plot the in-degree distribution graphs with a lognormal fit for each year and find the best $\sigma^2$ value to describe the tail. The $\sigma^2$ value associated with each year's lognormal fit is plotted against year ($\sigma^2$ plot) is shown in figure \ref{sigma2dvma}. For a good model we need $\sigma_{Model}^2(t) \approx \sigma_{data}^2(t) =1.8$ (from the data's $\sigma^2$ plot in green \ref{sigma2dvma}). Model A is not consistent with this with significantly lower value $\sigma_{ModelA}^2(t)\approx0.33 \pm 0.15 \ne1.78 \pm 0.14$ which also seem to show a downward trend as papers get older. In fact, we find that for all $p$ values ($0<p\le1$) the $\sigma^2$ plots from model A were very similar to figure \ref{sigma2dvma}.

%cat

Why is $\sigma^2 \approx0.33 \pm 0.15$ \textit{so small} for model A? This is because in our citation network the oldest papers (e.g.\ papers in years 0 and 1) will \emph{all} gain many citations through cumulative advantage (if $p\neq 0$) as they have been around the longest and will have had more chances to accumulate citations. In fact, our model shows that generally the oldest papers will be those in the fat-tail with many citations each.  In terms of the width $\sigma$ there is relatively little variation around the mean. Likewise the youngest papers (e.g.\ papers in years 9 and 10) will lose out and all will have a similarly low number of citations. The principle is the same for all years with relatively little variation around the means in the citation counts of papers published in the same year. Thus the Price model and the variation used for our model A is likely to have little variation and low $\sigma^2$ values for all years.

Our simple cumulative advantage model A with best parameter p=0.55 does produce an overall fat-tailed in-degree distribution for all years figure \ref{ModelAunall} (although not quite fat-tailed enough because model A's overall $\sigma^2 =1.05 \pm 0.18$, figure \ref{ModelAunall}, is not within the data's overall $\sigma^2 =1.78 \pm 0.14$, figure \ref{dataindeg}). It does produce a constant $\sigma^2$ plot, see previous section. However, model A does not produce the long-tail for \textit{individual} years. The average $\sigma^2$ with time for model A $0.33 \pm 0.15$ does not lie within the data's average $\sigma^2$ with time $1.78 \pm 0.14$. To increase the $\sigma^2$ values (over different years) we need a new parameter.

% ******************************************************************************
\section{Model B: Time Decay of Core Papers}\label{sec:MB}

%\subsection{Conclusions and Motivations for model B}\label{concA}

The problem with the Price model and our model A is that cumulative advantage gives the oldest papers too much of an advantage.  That is, all the papers in the oldest years tend to have a large citation count so that there is too little variation in their citation counts.  So for our second model, model B, we suppress the probability of adding a citation to an older paper.

\begin{figure}[h] %[H]
\begin{center}
% Note that the trim goes left, bottom, right, top
\includegraphics[scale=0.4]{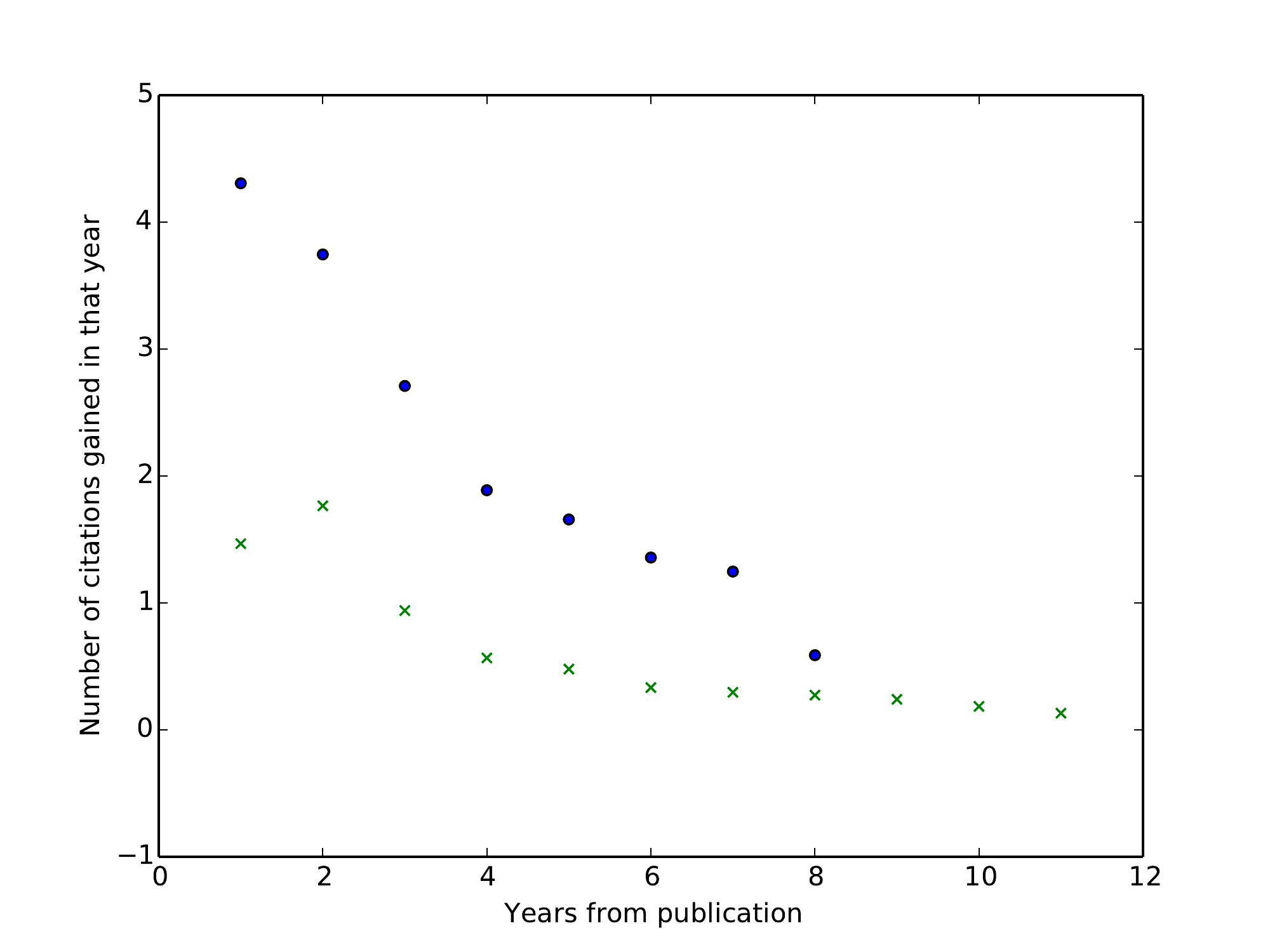}
\end{center}
\caption{The average number of citations gained each year for hep-th arXiv papers published in 1992 and 1995, shown as crosses and dots respectively. The horizontal axis gives the number of years since publication, the vertical axis is the number of citations gained in that one year. The first two years of the data, 1992 and 1993, have anomalous distributions with a peak in year 2, see section \ref{NPPPY}. For later years, e.g. 1995 onwards, we find a sharp decrease which may be characterised using an exponential decay.  The average number of citations accumulated in one year falls by around a half over 3.5 years or around 5,000 papers, figure \ref{fig:data_papers_per_year}. }
%Note if you want to reference in the caption \protect{\cite{citationlabel}}.
\label{exp1} \label{exp295}
\end{figure}

Looking at our data, figure \ref{exp1} shows how citations are gained over time for papers published in the first (1992) and fourth (1995) years of our hep-th data. This shows a general decay in the rate at which citations are accumulated, also found in \cite{ISI:000088918500004}.
In order to keep our model simple we will use an exponential decay form in model B, giving it an additional parameter over model A. Such exponential decays are widely used in citation modelling \cite{ISI:000187183600031,ISI:000302837300019,ISI:000225682200007,ISI:000088960600044,ISI:000168730800034,ISI:000165399000056,ISI:000295265100032}. Alternatives, such as only referencing papers if they are less than a year old, also creates an effective time decay \cite{AMTC}. A similar decay over time in attachment probability has been used for other types of citation networks such as patents \cite{ISI:000222600600002}.

In our model B we add new publications one at a time as before.  With probability $p$ the new publication chooses to reference an existing paper $i$ chosen with a probability proportional to the current in-degree of paper $i$ multiplied by an exponential decay factor dependent on $i$'s age. Alternatively, with probability $(1-p)$ a papers are chosen with uniform probability multiplied by the same exponential decay factor. Thus the probability of attaching to node $i$, $\Pi_B(i)$, is roughly\footnote{Again there is a small correction to this form to allow for the fact that we do not allow the same vertex $i$ to be chosen more than in model B.}
\begin{equation}\label{AttachProbB}
 \Pi_B(i)
 =
 p    \left( \frac{\kin_i 2^{(N-i)/\thalf } }{Z_{\mathrm{B,ca}}}\right)
 +
 (1-p)\left( \frac{2^{(N-i)/\thalf } }{Z_{\mathrm{B,ua}}}\right) \, ,
\end{equation}
where $\kin_i$ is the in-degree of node $i$, $\thalf$ is defined as the `attention span' parameter in the \textit{model}\footnote{This is different from the `half-life' values referred to later, which are \textit{measured} from the data.}, this is a time decay parameter in the model that acts like the time it takes for a paper to gain half the total number of citations it ever will, when $p=0$). $N$ is the total number of pre-existing nodes a publication can reference. Note that we are using the `rank' time to determine the age of the paper, that is paper $i$ is added at a time equal to $i$. The normalisation factors, $Z_{\mathrm{B,ca}}$ and $Z_{\mathrm{B,ua}}$, are
\bea
Z_{\mathrm{B,ca}} &=& \sum_{j=1}^{j=N} k_j^{\mathrm{in}} 2^{-(N-j)/\thalf}.
\label{ZBca}
\\
Z_{\mathrm{B,ua}} &=& \sum_{j=1}^{j=N} 2^{-(N-j)/\thalf} = \frac{1-2^{-N/\thalf}}{1-2^{-1/\thalf}}.
\label{ZBua}
\eea

\begin{figure}[h] %[H]
\centering
\includegraphics[scale=0.4]{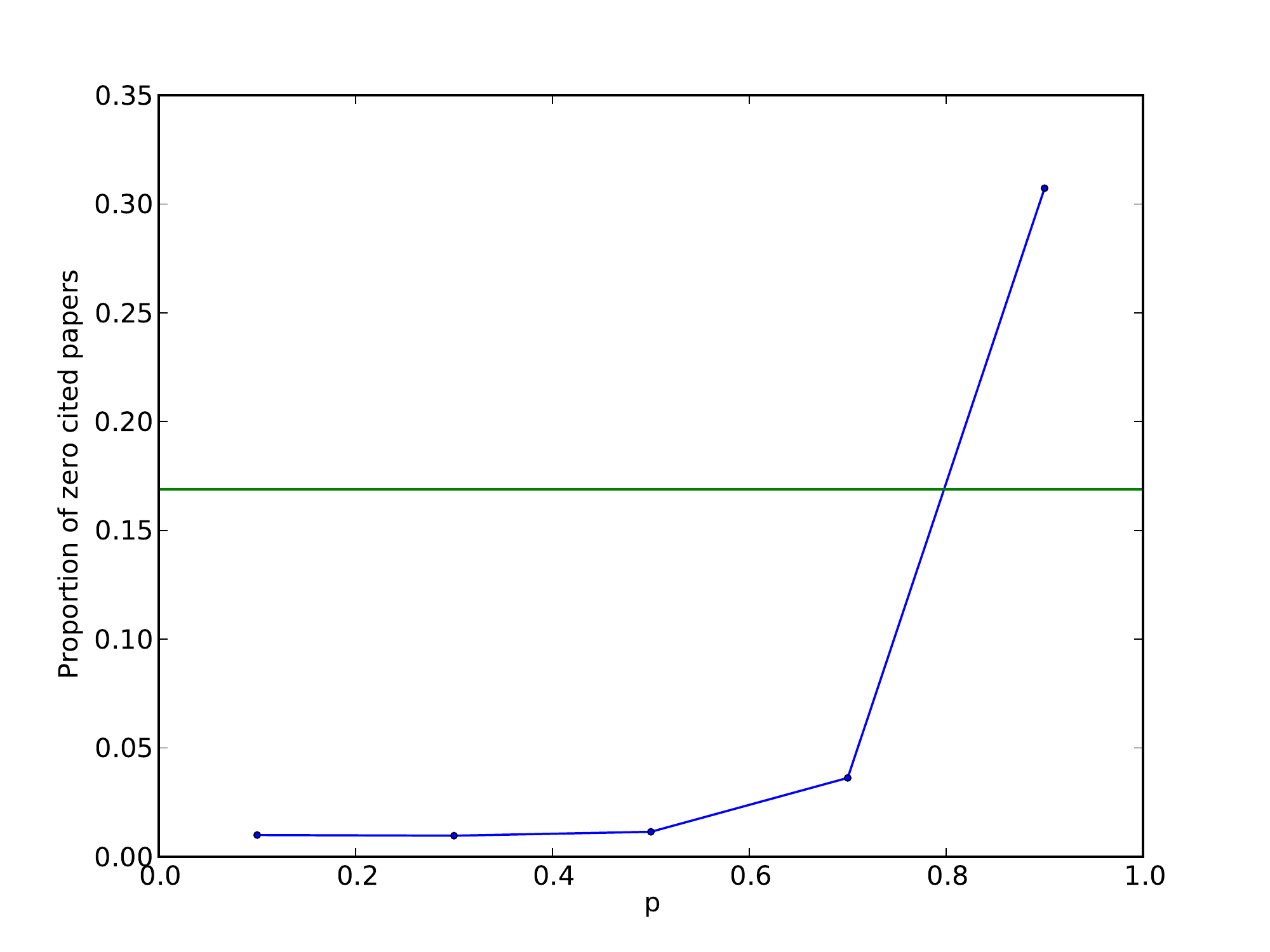}
\caption{The blue data points show the $z$ found for model B for a fixed $\thalf$  of 2000 papers and given probability of cumulative advantage $p$, against varying $p$. The green horizontal line gives the value for $z$ found in the hep-th data, namely  $z=0.169$. Interpolating form the model output suggests a value of around $p=0.80$ is needed.}
\label{fig:best_p_value_modelB}
\end{figure}

Thus our model B has just two parameters: $p$ and $\thalf$. To determine $p$ we use the fraction of zero cited papers found in the whole of our hep-th data, $z=0.169$. With a fixed attention span we vary $p$ and calculate $z$ for each of these models. We worked with a value of $\thalf=2000$ paper while determining $p$, see figure \ref{fig:best_p_value_modelB}. From this we find $p=0.80$. We found that this result for $p$ does not change significantly for different attention span values $\thalf$. For example, for attention span values of $\thalf=200$ and $\thalf=5000$ papers we need to choose $p=0.81$ and $p=0.79$ respectively to get the correct zero-cited paper fraction, a mere $1.25$\% change from our chosen value of $p=0.80$.

Given $p=0.80$ we now wish to determine the best half-life value for model B. To do this we choose the value of the attention span parameter $\thalf$ such that \textit{measured} half-lives of the hep-th data are as close as possible to the \textit{measured} attention span outputted by the model. This process of trying to match the half-life derived from the output of a citation network model to that observed in data is original. However, there is no guarantee that the measured attention span parameter from our model will be identical to the half-life $\Thalf$ measured from the decay in citations seen in the data. This is because our model has two parameters $\thalf$ and $p$, both of which are dependent on time. $\thalf$ and $p$ cause recent papers to be more and less likely to be referenced, respectively, therefore the \textit{measured} attention span from the model is not equal to the inputted attention span $\thalf$.

\begin{figure}[h]
%        \centering
%        \begin{subfigure}[b]{0.485\textwidth}
%                \centering
%                \includegraphics[scale=0.44]{halflife_histogram1992.pdf}
%                \caption{This is a histogram of the number of papers published in 1992 with a given median half-life against median half-lives, both measured in calendar years. ??? CHANGE HORIZONTAL AXIS LABELS TO \textbf{yyyy} FORMAT ???}
%                \label{hldataearly}
%        \end{subfigure}
%        ~
%        \begin{subfigure}[b]{0.485\textwidth}
%                \centering
%                \includegraphics[scale=0.44]{halflife_histogram2001.pdf}
%                \caption{This is a histogram of the number of papers published in 2001 with a given median half-life against median half-lives, both measured in calendar years. ??? CHANGE HORIZONTAL AXIS LABELS TO \textbf{yyyy} FORMAT ???}
%                \label{hldatalate}
%        \end{subfigure}
        \begin{center}
                \includegraphics[scale=0.44]{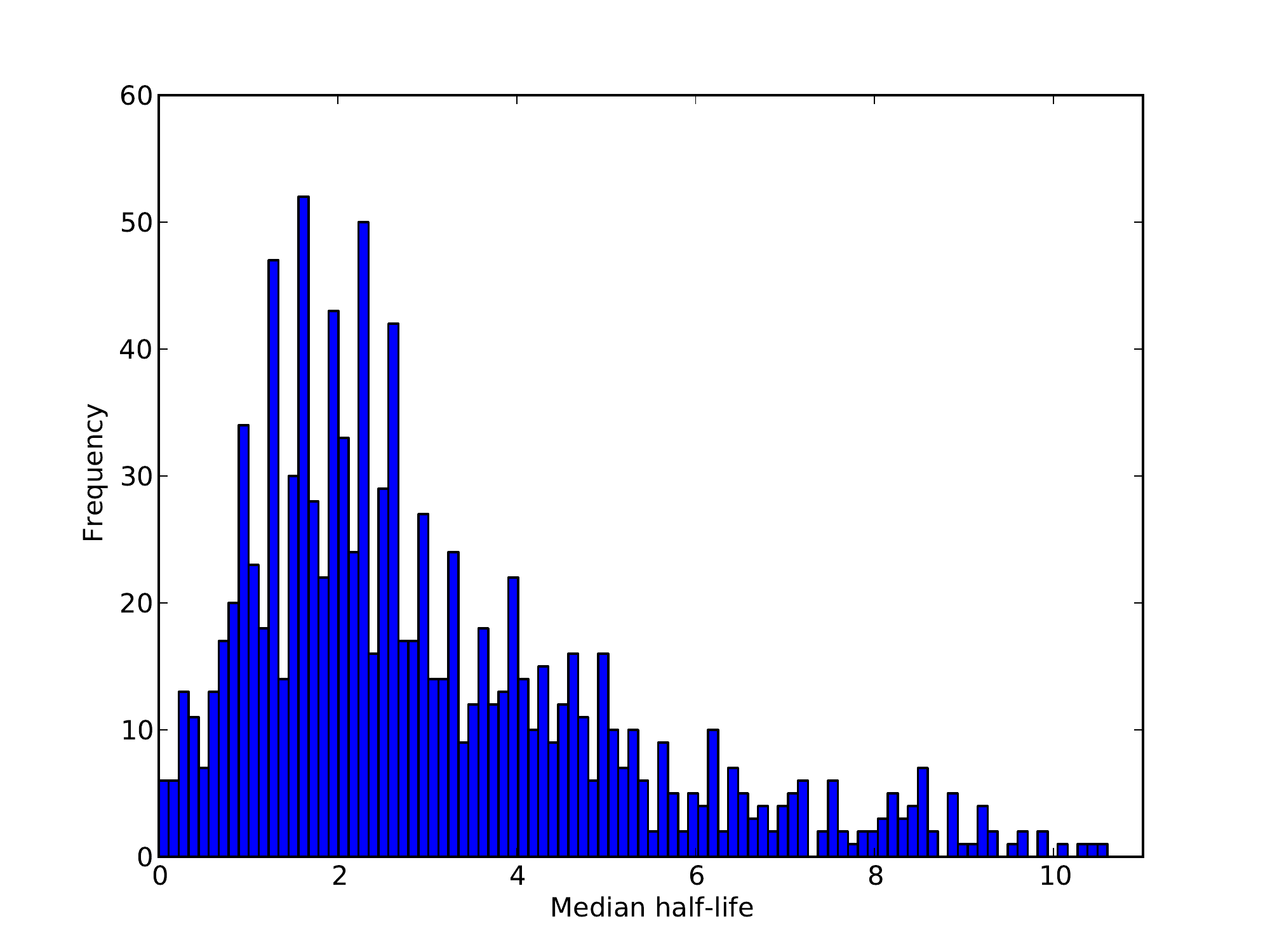}
                \label{hldataearly}
                \includegraphics[scale=0.44]{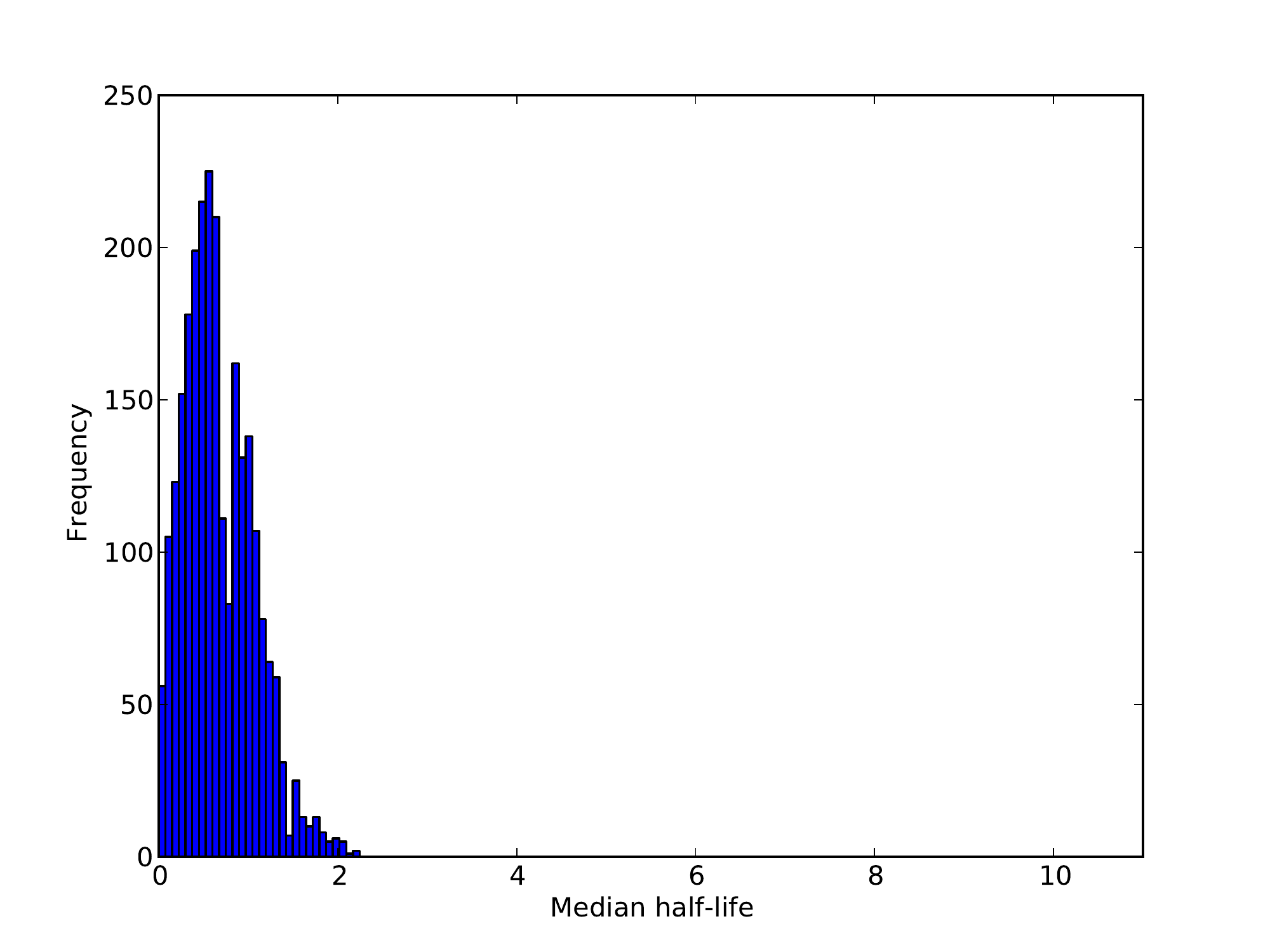}
                \label{hldatalate}
        \end{center}
        \caption{These are two histograms to contrast. Plotted are the number of publications created in year $y$ with a given median half-life $\Tmed$ against the median half-live value for an early year, above, $y=1992$, and a late year, below, $y=2001$. Both histograms have the same $\Thalf$ bins.  They are similar because they both follow an approximate skew normal distribution with a mean of 2 and 0.5 years for 1992 and 2001 respectively.
        %The effect of different time scales for the acquisition of their citation sis clearly visible Also, the width is low (and the peak is high) for 2001 and the width is higher (peak is lower) for 1992 because 2001 only had a maximum of 2 years to gain citations, whereas 1992 has had a maximum of 11 years. Because 1992 has longer to gain citations it also has a higher median half-life. Given these curves are different, how do we determine the \textit{half-life} we should input into model B? All years here are calendar years.
        }
        \label{hldata}
\end{figure}

Our approach was to take each paper in hep-th and to find the median citation time $\Tmed$, that is the time it took for that paper to accumulate half of its final citation count (after 11 years). The distribution of these median times were considered for papers published in the same year, with examples of these distributions shown in figure \ref{hldata}.
There are large fluctuations in the observed median times, but, we will just use the average median time for each year $y$ in our data set, $\texpect{\Tmed}_y$, to characterise the actual decay over time seen in our citation data. For an exponential model, the total number of citations in our model over a period $T$ is proportional to
$\int_{0}^{T} dt\;2^{-t/\Thalf } = (1-2^{-T/\Thalf}) (\Thalf / \ln(2))$. Since the number of citations at the median time $\Tmed$ is half that of the total time $\Ttotal$ available for a paper to collect citations in a given data set, we have that
\begin{eqnarray}
2^{1-\Tmed/\Thalf} - 2^{-\Ttotal/\Thalf} &=& 1 \, .
\label{nonanal}
\end{eqnarray}
Using the average median time $\texpect{\Tmed}_y$ for papers published in one year of our hep-th data we solve \tref{nonanal} numerically to find a data half-life value $\Thalf$.

%\tnote{DONE (T).  (S) please check the above equations one more time. I'm sure they're fine but it's good to check.}

\begin{figure}[htbp]
        \centering
        \begin{subfigure}[b]{0.45\textwidth}
                \centering
                \includegraphics[scale=0.38]{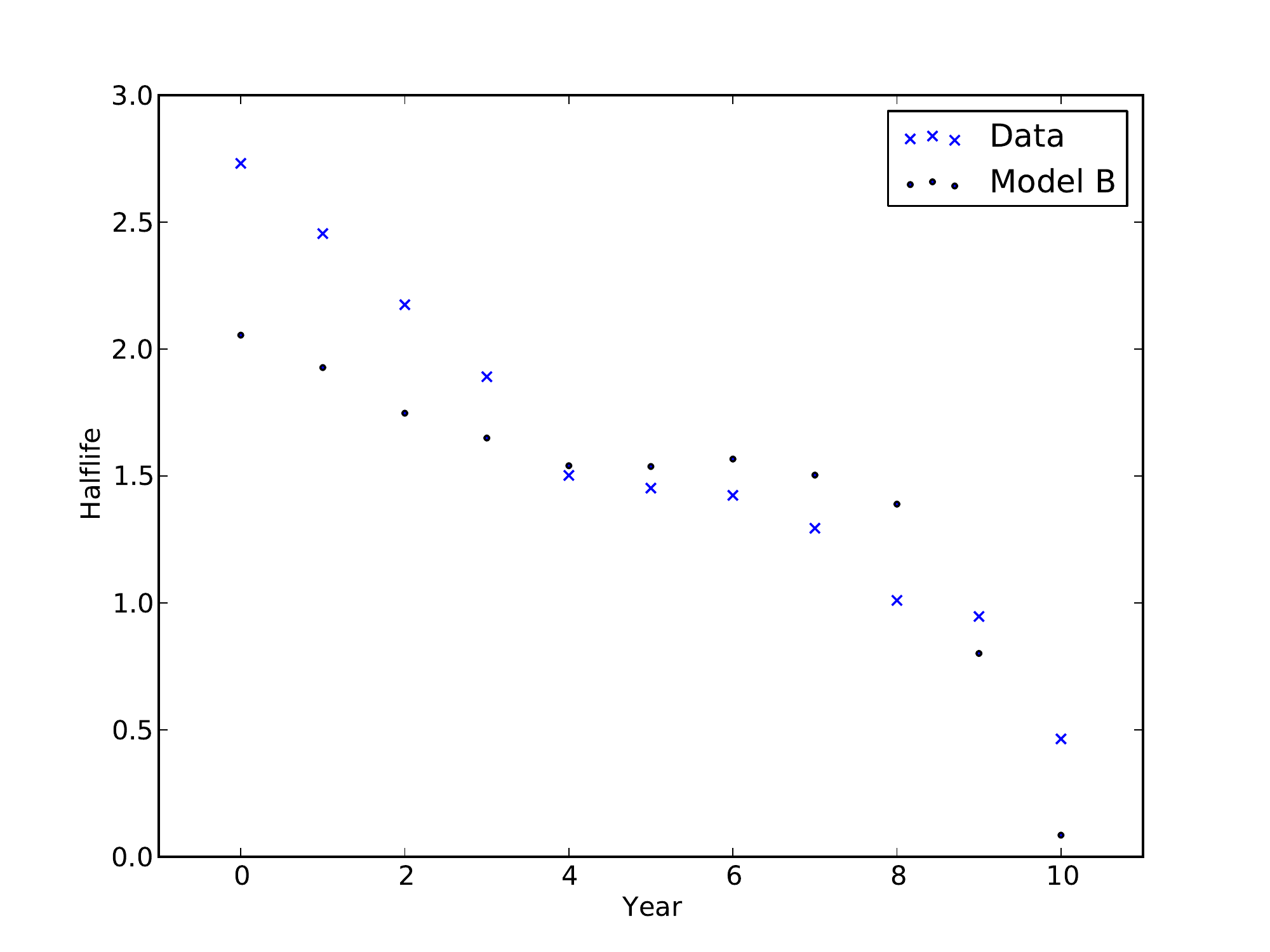}
                \caption{This is a plot of outputted half-lives for the hep-th data, crosses, and final model B, dots, (for p=0.80, $\thalf=2000$ papers) against calendar year. Years 1992, 1993 etc. are relabelled as years 0, 1 etc. }
                \label{Bhl_normal}
        \end{subfigure}
        ~
        \begin{subfigure}[b]{0.45\textwidth}
                \centering
                \includegraphics[scale=0.38]{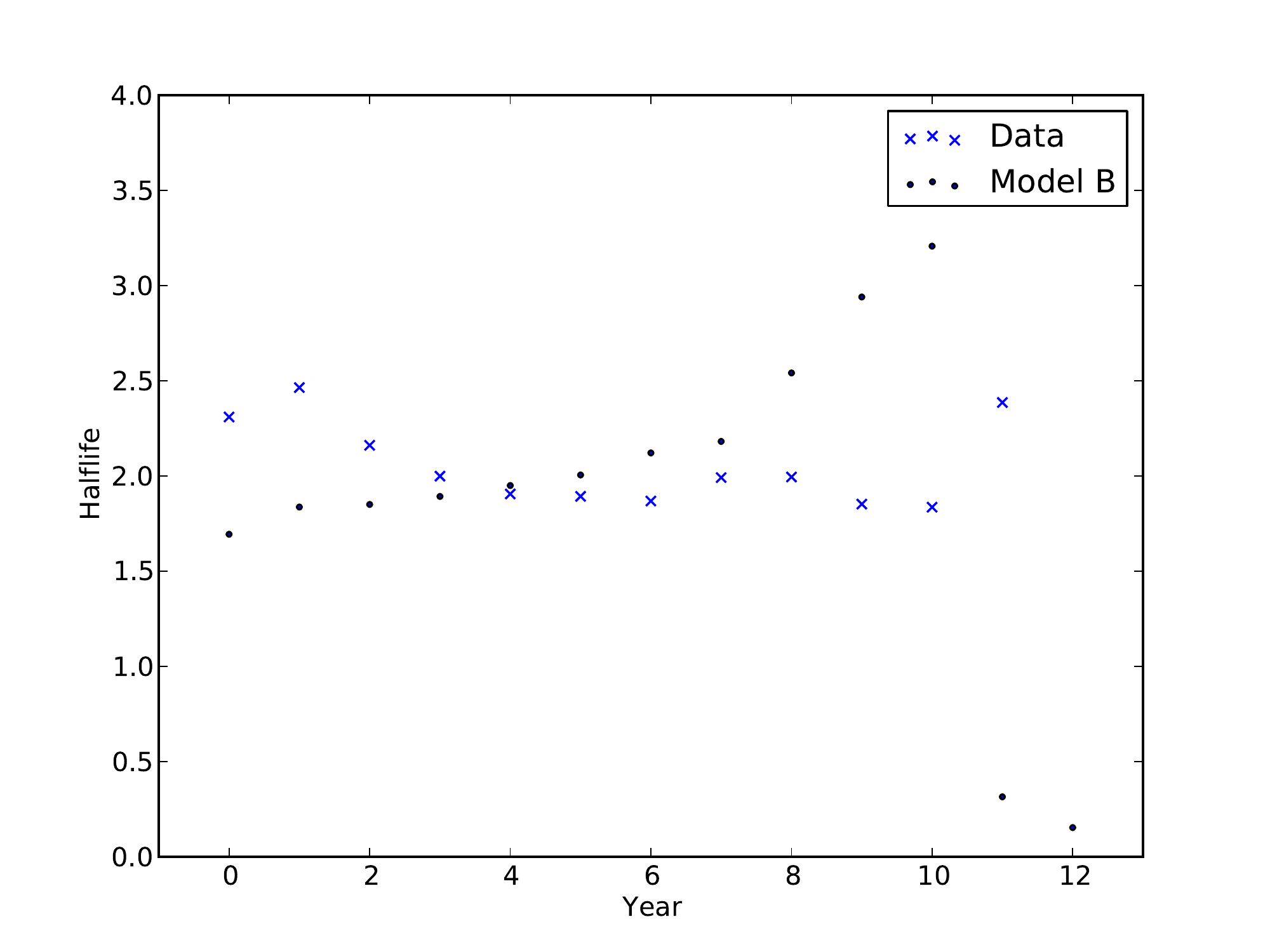}
                \caption{This is a plot of outputted half-lives, measured in rank time where a year is 2000 papers for the data (crosses) and final model B output with $p=0.80$ and $\thalf=2000$ papers (dots). Year 0 is the 1st 2000 papers, etc. created in the network.}
                \label{Bhl_rank}
        \end{subfigure}
        ~
        \caption{In figures \ref{Bhl_normal} and \ref{Bhl_rank} we compare/contrast the measured half-life $\Thalf$ for the hep-th data (crosses) and the output from model B using the best parameter values of $p=0.80$ and $\thalf=2000$ (dots), for normal time (left) and rank time (right) in years.}
        \label{Bhl}
\end{figure}

Figure \ref{Bhl_normal} shows that the half-lives characterising the data, $\Thalfdata$, decrease as the for later years. This is because the number of papers published per year is increasing and your citations per year decrease if there is more literature to read. To factor this growth rate out, we can work in `rank time'. That is, since arXiv automatically provides the order in which papers were first submission, we use this as a time parameter, so the earliest paper in the data has rank time 1, the $n$-th paper submitted is given rank time $n$.  In this case we define our `years' to be collections of 2000 papers, so year 0 contains the first 2000 papers, year 1 are papers with rank times 2001 to 4000, and so forth. This gives us the roughly the same number of years as the calendar years in our data. Working out the data half-life $\Thalfdata$ using rank time we find this is now roughly constant, see figure \ref{Bhl_rank}, confirming our suggestion that the number of papers in each year is an important factor here.

\begin{figure}[h] %[H]
\centering
\includegraphics[scale=0.5]{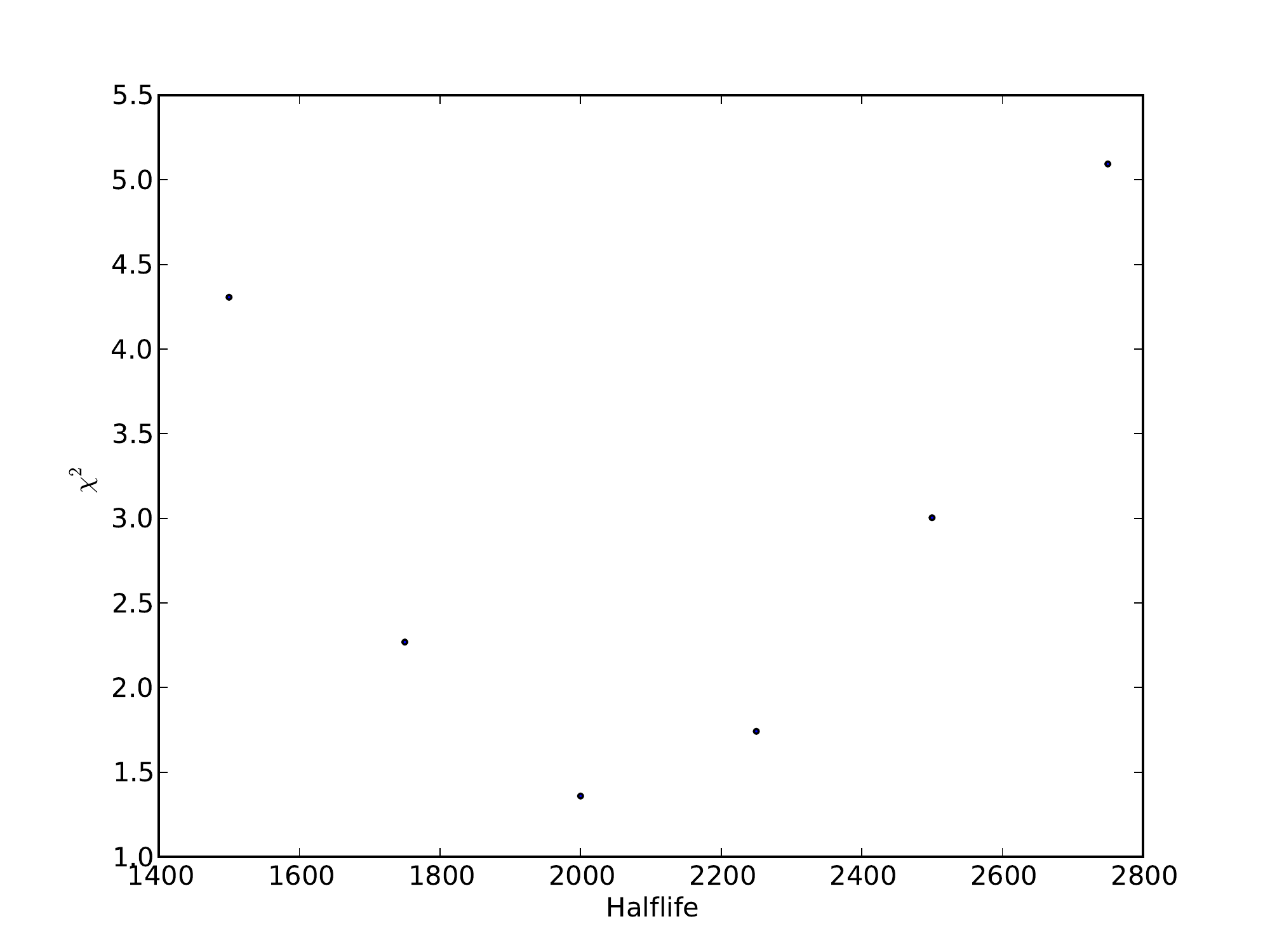}
\caption{This is a plot of $\chi^2$, see equation \ref{eq:Bchi2}, against varying input parameter $\thalf$, measured in rank time (number of papers), constant $p=0.80$ in model B. There is a clear minima, minimum difference, between model B and the hep-th data when $\thalf=2000$ papers. Therefore a half-life of 2000 papers is chosen as our final $\thalf$. The minima approaches zero but is non-zero, meaning the half-life v time plot for the best model B is close to but not exactly equal to the data, figure \ref{Bhl}.}
\label{Bchi2}
\end{figure}

In order to set the model parameter $\thalf$ we now try different values of $\thalf$ for model B and use the output from the model to determine a model output half-life $\Thalfmodel$ in exactly the same was as we did for the arXiv data, working now in rank time. To find the best $\thalf$ parameter value we minimised $\chi^2$ where
\begin{eqnarray}\label{eq:Bchi2}
\chi^2(\thalf, p=0.80)= \sum_{y} \left(\Thalfdata^{(y)} - \Thalfmodel^{(y)}(\thalf, p=0.80) \right)^2 \, .
\end{eqnarray}
and $\Thalfdata^y$ and $\Thalfmodel^y$ (calculated above) are the average half-lives outputted of the data and model B (for $p=0.80$ and given $\thalf$) for rank year $y$, respectively. Figure \ref{Bchi2} shows the results with a minimum $\chi^2$ found for $\thalf=2000$ papers.

As a final check, Figure \ref{Bhl} shows the half-lives $\Thalf$ found from the arXiv data and from the output of model B with its optimal parameter values $p=0.80$, $\thalf=2000$, both in normal time and rank time. The results are not perfect but we believe results are acceptable for a simple two parameter model.  We also note our earlier assertion that the internal half-life parameter $\thalf$, which is set here to 2000 papers, does not need to match the measured half-life value in the output.  For instance, using rank time, the measured half-lives are around 1.5 to 2.0 years, so 3000 to 4000 papers. As explained above, cumulative advantage favours older papers which extends $\thalf$ measured (outputted).

% ------------------------------------------
\subsection{Results for model B}

\begin{figure}[h] %[H]
\centering
\includegraphics[trim = 0cm 0.5cm 0cm 1.5cm, clip = true, scale=0.5]{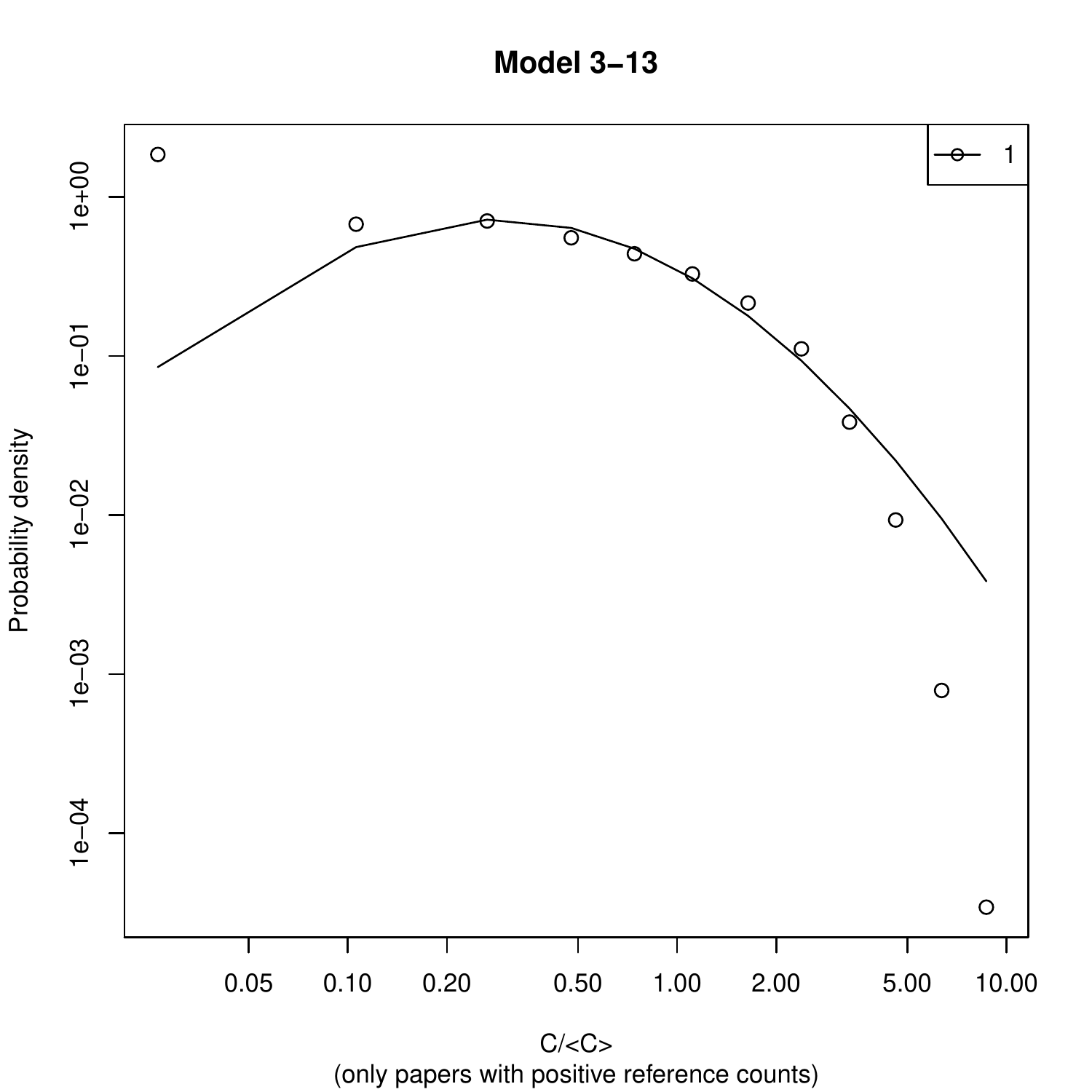}
\caption{This is the overall in-degree distribution of probability density against normalised citation count for model B with best parameters $p=0.80$ and $\thalf=2000$ papers. This $\sigma^2$ of this distribution is $1.05\pm 0.18$. Compared to model A, figure \ref{indegma}, model B has more papers with high citation counts (the mid part of the distribution) making the tail longer and therefore increasing $\sigma^2$ (although the distribution maximum only goes up to approximately 10 citations like model A, figure \ref{indegma}). However, compared to the data, figure \ref{dataindeg}, the tail of model B needs to be longer, it needs to have a non-zero probability of having a normalised citation count greater than 50.}
\label{indegMB}
\end{figure}

We can now look at the shape of the citation distribution for model B with the optimal parameter values, $p=0.80$ and $\thalf=2000$ papers. We plot the in-degree distribution for the whole data set and observe the characteristic large width, figure \ref{indegma} as expected by the hep-th data, figure \ref{dataindeg}. The width of the in-degree distribution is a slight improvement from that found with model A, changing from $\sigma^2 =1.05\pm 0.18$ in model A to $\sigma^2 =1.14 \pm 0.17$ in model B, closer to the arXiv data's $\sigma^2 =1.78 \pm 0.14$. However, as our error estimates show, this is not \textit{significantly} better statistically. In fact we tried varying the parameters $p$ or $\thalf$ but found no way to increase the $\sigma^2$ significantly in model B. See figure \ref{indegMB} for the in-degree of all years for model B.

\begin{figure}[h] %[H]
\centering
\includegraphics[width=0.5\textwidth]{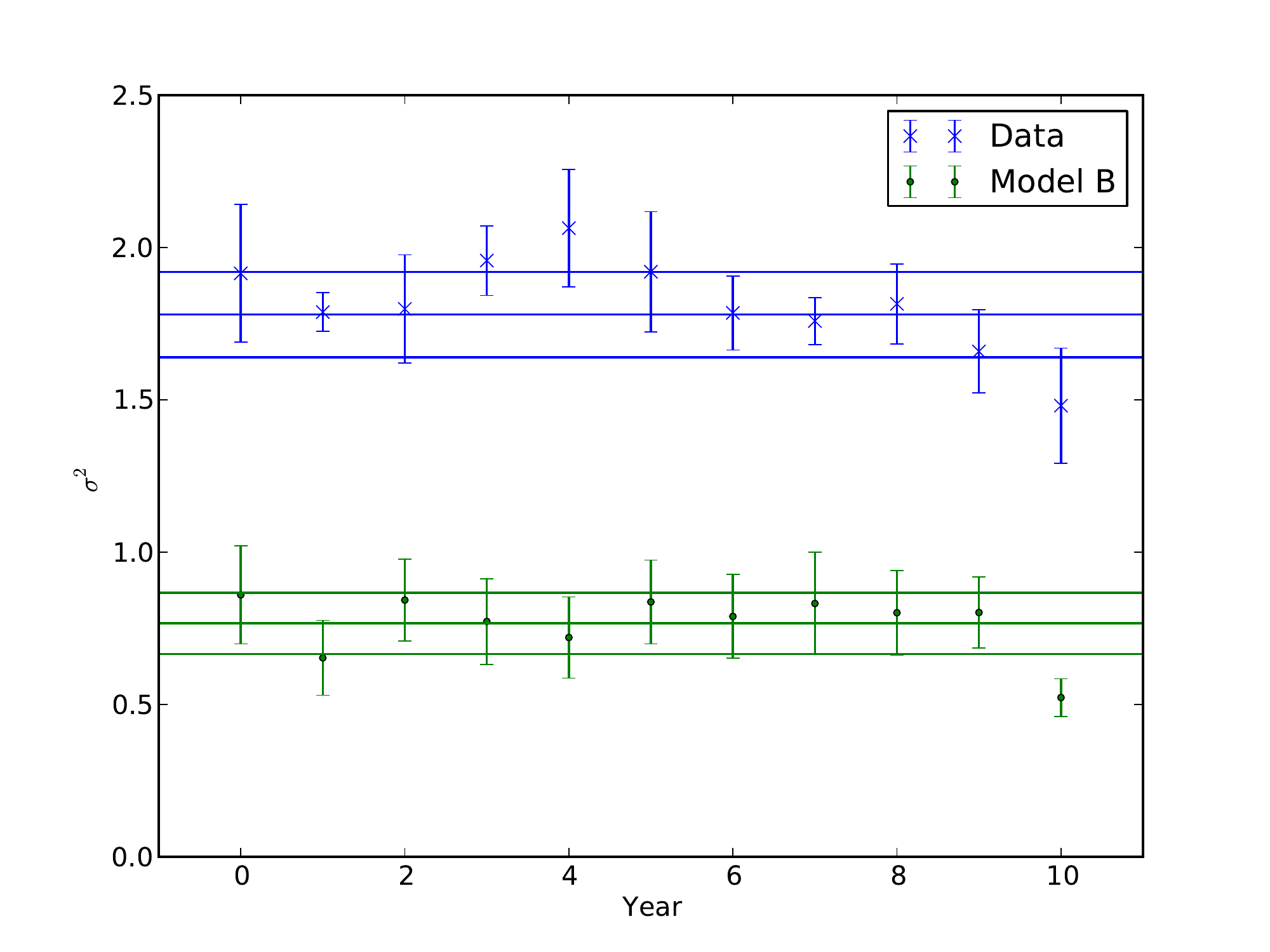}
\caption{This is a $\sigma^2$ plot of the data, blue crosses, and model B for $p=0.80$ $\thalf=2000$ papers, green dots. The average, upper and lower bounds (one standard deviation away form the mean) of the data and model B at $1.78\pm 0.14$ and $0.77 \pm 0.10$ in blue and green, respectively. We observe model B is not a good fit of the data as the data's average $\sigma^2$ is approximately 2.3 times the corresponding average for model B. No variation in the parameters $p$ or $\thalf$  altered the $\sigma^2$ plot of model B significantly. We can conclude from this graph model B is not a good citation network model of the hep-th data. The signature width is not large enough, however, it is constant.}
\label{Bsigma2}
\end{figure}

For model B, the width of the citation distribution for papers in each year with its optimal parameter values, $p=0.80$ and $\thalf=2000$ papers, are shown in figure \ref{Bsigma2}. We find that, like the arXiv data, model B is now producing a distribution with a roughly constant width $\sigma^2$, an improvement over the decreasing width with age found in model A.  Unfortunately model B is still producing a width that is about half that seen in the data. Although the $\sigma^2$ plot for model B, averaging at $0.77 \pm 0.10$, is closer to the hep-th data $1.78\pm 0.14$ than model A, averaging at $0.33 \pm 0.15$, no amount of varying the parameters $p$ or $\thalf$  increased the $\sigma^2$ plot significantly, i.e.\ within each others error bars. Therefore we need to add a further parameter to model B in order to reach wider range of citations in each year, and so as larger $\sigma^2$.

% ******************************************************************************
\section{Model C: Copying}\label{sec:MC}

A general problem with the Price model and related models, such as our models A and B, is that the cumulative advantage and random attachment processes require global knowledge of the whole network.  This can be seen in the normalisation of the two contributions in $\Pi_A$ of \tref{AttachProbA}: the number of citations for the cumulative advantage process and the number of papers for the uniform random attachment process.  Neither of these is needed or known by authors looking to cite new papers.  The addition of a time decay in model B means that the emphasis is on a smaller set of recent papers, something authors are more likely to be aware of, but the normalisations in $\Pi_B$ \tref{AttachProbB} indicate that authors still require global knowledge if the processes are taken literally. There is, however, a very natural process based on local knowledge which reproduces the long tails and that is to use random walks \cite{AMTC, Vaz00,KR01,Vaz03,FLDG03,SK04,ES05}.  That is an author will find papers of interest by following the references of papers already known to the author.  In terms of a model, we will assume that authors find new papers of interest by choosing uniformly at random from the references listed in one paper; authors are making a random walk on the citation network.  To do this, an author need only use the the local information available in currently known papers.  The properties of the rest of the network are irrelevant to this process as indeed they will be for real authors. What is particularly interesting, is that random walks of any length or type, even of length one, will generate the fat-tailed power-law like distributions with a wide range of characteristics \cite{ES05}.

In looking to improve upon model B, we will therefore add a local search process based on a single step random walk to find some of the citations to be added to a new paper. We will start these walks from papers chosen (and cited) using the same mechanism as model B, given the partial success obtained there.   That is we are assuming that some citations are derived using global knowledge and some from local searches.  This is meant to mimic the fact that authors do come to a new paper with some limited knowledge of the whole network, obtained by looking at recent posts on arXiv or from conversations with colleagues and so forth, while local searches will also reveal new relevant material to an author.

\begin{figure}[h] %[H]
\begin{center}
\includegraphics[width=0.4\textwidth]{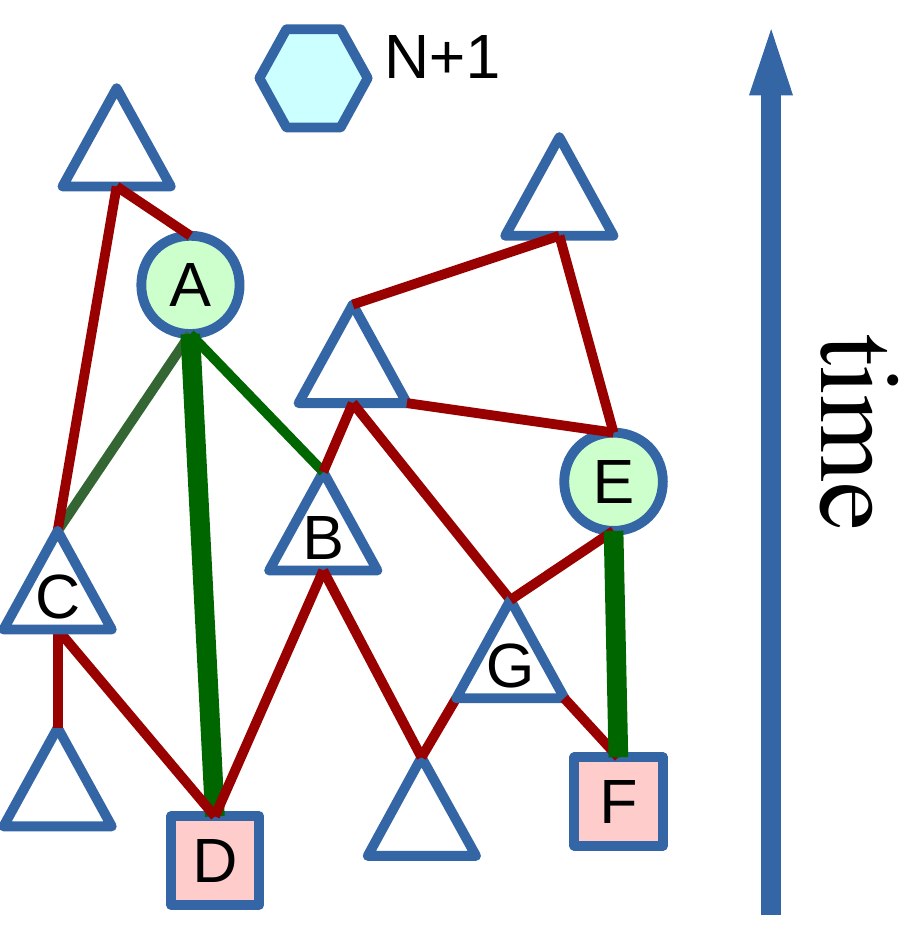}
\end{center}
\caption{An illustration of the processes in model C. A new paper $(N+1)$, the blue hexagon, is added to the network. Then the length of the bibliography $r$ is chosen from a normal distribution of the same mean and standard deviation as the arXiv hep-th data.  Suppose $r=4$ here. A first `core' paper, paper A (green circle), is then chosen exactly as in model B, using global information from all previously published papers (triangles, circles, dots and squares).  Next, all the papers cited by paper A, are considered and each is added to the bibliography of the new paper $(N+1)$ with probability $q$. These are our `subsidiary' papers in the reference list of the new paper $(N+1)$. In this case paper $D$ (red square) is chosen as indicated by the thick green link while the thin green links indicate that neither paper B nor C (white triangles) are chosen this way.  Now a second core paper is chosen via global information (model B), say paper E (green circle), and added to the bibliography of the new paper $(N+1)$.  We again consider the papers cited by paper E, selecting them with probability $q$.  Here paper F is considered first and is selected becoming our second subsidiary paper citation.  At this point we have the four papers needed for the new paper so we do not consider paper G. The new paper cites two core papers, A and E (circles), and two subsidiary papers, D and F (squares).  The new citation network shown in figure~\ref{Modelcnew}.}
%Note if you want to reference in the caption \protect{\cite{citationlabel}}.
\label{Modelc}
\end{figure}

Model C is defined as follows, see figure \ref{Modelc}.  A new paper, paper number $(N+1)$, is added to the network and the length of its bibliography, $r$ is chosen randomly via a normal distribution as before. Next, we use the model B processes, $\Pi_B$ of \tref{AttachProbB}, to find the first `core' paper to be cited (through global knowledge of the entire network). We then look at \emph{each} of the papers cited by this first core paper, adding each of these `subsidiary papers' to the bibliography of the new paper $(N+1)$ with probability $q$ (through local knowledge of the bibliography of this core paper). We will then repeat this whole process, choosing another `core' paper $c'$ using the global knowledge processes of model B, followed by more `subsidiary' papers found using the local knowledge process of a one-step random walk from the new core paper. Papers will only be cited once and the whole process stops as soon as the new paper $(N+1)$ has a total of $r$ references. For more on declustering refer to \cite{GoldC4}.

Note that the time scale used when choosing the core papers using global information, the attention span $\tau$, is applied only to the selection of core papers in model C.  This is to be contrasted with model B where all papers cited are core papers selected using global information with one time scale $\tau$. There are many possible variations of our model C but we expect that similar results can be obtained as suggested in \cite{ES05}. We prefer to keep the model as simple as possible. This leaves us with just three parameters for our model C: $p$ and $\AS$ from model B and the additional probability $q$.
%Refer to the appendix for the probability of a new node in model C attaching to a pre-existing node \ref{ApModC}.

\begin{figure}[h] %[H]
\centering
\includegraphics[width=0.48\textwidth]{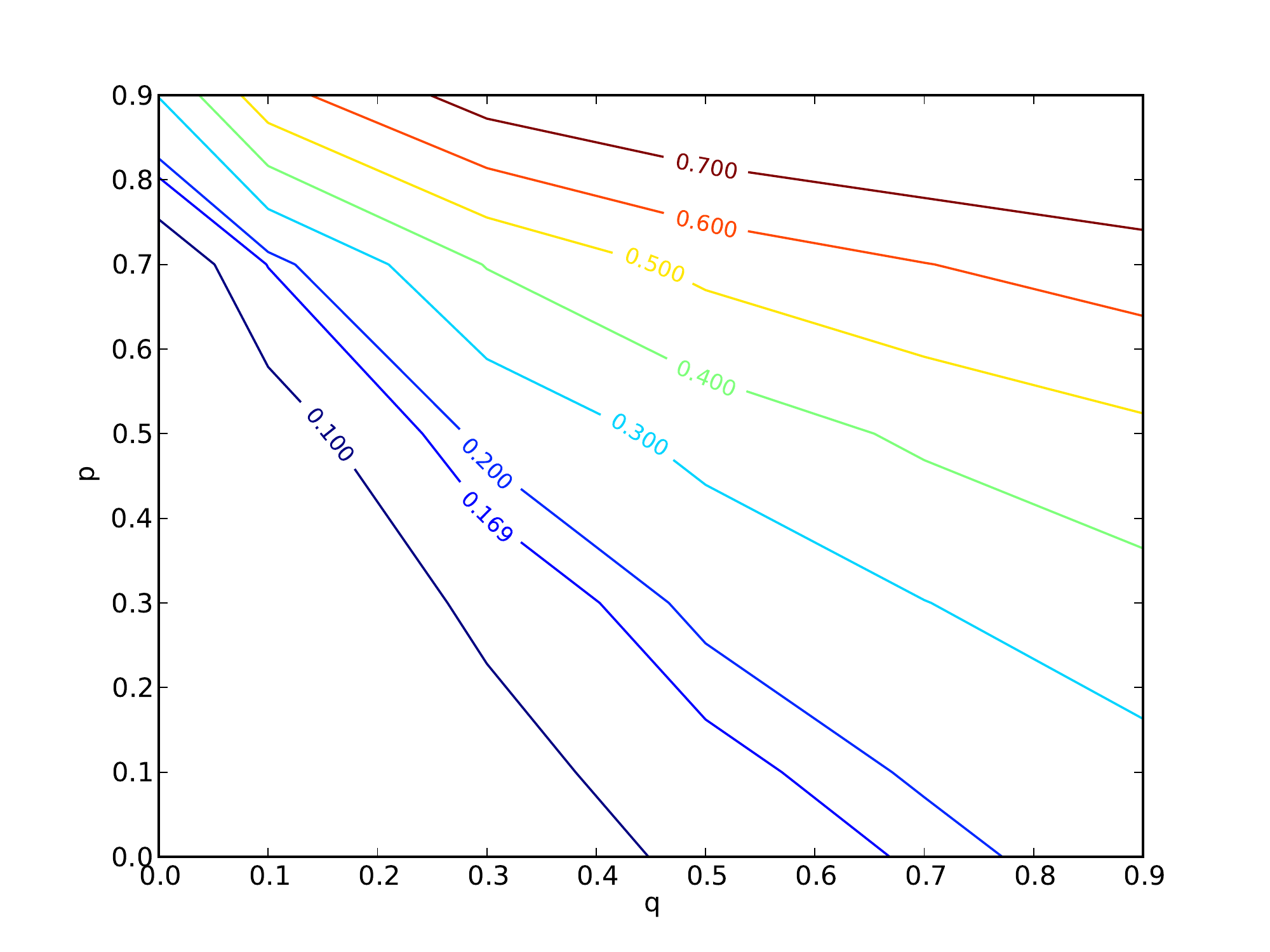}
\includegraphics[width=0.48\textwidth]{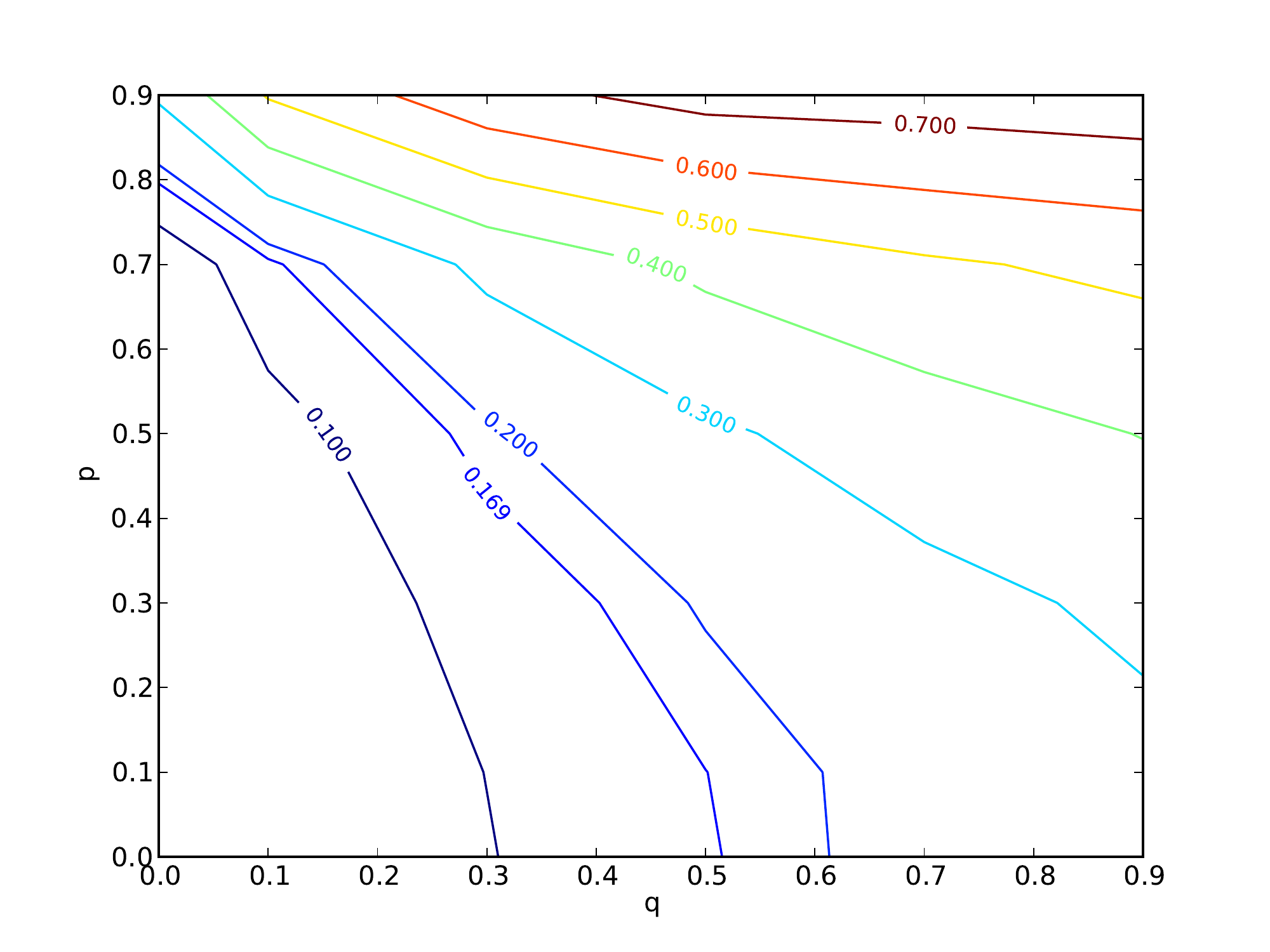}
\caption{The fraction of zero cited papers found in the output from model C. The lines represent different fractions of zero cited papers $z_c$ found for varying $(p,q)$ values for a fixed attention span (defined below) of $\AS=200$ papers on the left and $\AS=2000$ papers on the right. The line of constant $z_c$ equivalent to the hep-th data's $z$ value, $z_\mathrm{data}=0.169$. There are a family of $(p,q)$  solutions ranging from $(p,q)=(0.8, 0)$ to $(p,q)$$=(0,0.45)$, one of the purple contour lines. Note that $(p,q)=(0.8, 0)$, is simply model B, i.e.\ there are no $q$  steps only a mixture of cumulative advantage and uniform attachment with an overall time-decay factor. For the `unreasonable' value of $\AS=2000$ on the right, the solution of $(p,q)=(0.0,0.80)$ for $z_\mathrm{modelC}=z_\mathrm{data}=0.169$ is the same as in contour for $\AS=200$ papers. Along the $q$ axis the solution for $z_\mathrm{data}=0.169$ when  $\AS=2000$ is $(p,q)=(0,0.51)$, as opposed to $(0,0.66)$ for $\AS=200$. Although the attention span $\AS$ has changed by a factor of 10, $q$  has only changed by $0.15$. For $q<0.3$ this contour can be treated as approximately equivalent to that for $\AS$ =2000 papers. Therefore it is valid to just use one contour plot for $\AS=200$ papers. }
\label{contour}
\end{figure}

To fix the unknown parameters we start by using the constraint that the total fraction of zero cited papers found in the output of model C, $z_c$, should be equal to that found in the data, $z=0.169$. To find these optimal values in figure \ref{contour} we show contours indicating the fraction of zero cited papers $z_c$ in the output of model C for a given $p$ and $q$ with the third parameter $\AS$ fixed. By looking at the widths of the citation degree distributions $\sigma^2$ against year in the model output, we found that `reasonable' attention spans lay between 150 and 300 papers.  Further in this range of $\AS$ the contour plots of $z_c$ for different $(p,q)$ values did not change significantly. Even when we tried much larger values for $\AS$, we found that the contour plots for $z_c$ against $(p,q)$ were approximately the same for $\AS$ =2000 and 200 papers if $q$  was less than about 0.3. In fact we find that $q=0.2$ is our chosen value suggesting that $z$ provides only a weak constraint on $\AS$ in the region of interest. However we will take the attention span to be $\AS=200$ papers, partly because it works for all values of $(p,q)$ and doesn't bias the determination of the best $(p,q)$ values.

To fix $p$ and $q$ we need a further piece of information.  To do this we conjectured that the value of $q$ required will depend on the fraction of references to core papers in each paper. In the model C there will be a number of references made from new publications of which around $q \texpect{\kout} =q \texpect{\kin}$ go to subsidiary papers
for every citation to a core paper. So a low $q$ will require a large faction of core papers to make up the bibliography, and vice versa.

\begin{figure}[h] %[H]
\begin{center}
\includegraphics[width=0.45\textwidth]{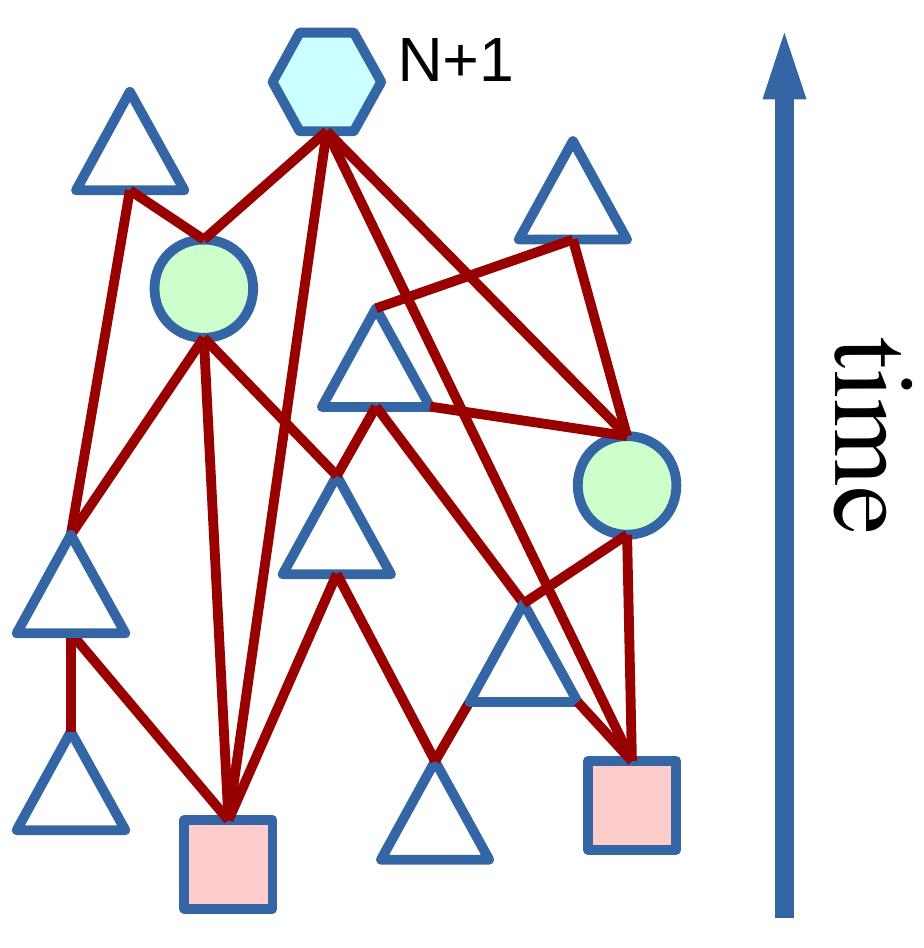}
\includegraphics[width=0.45\textwidth]{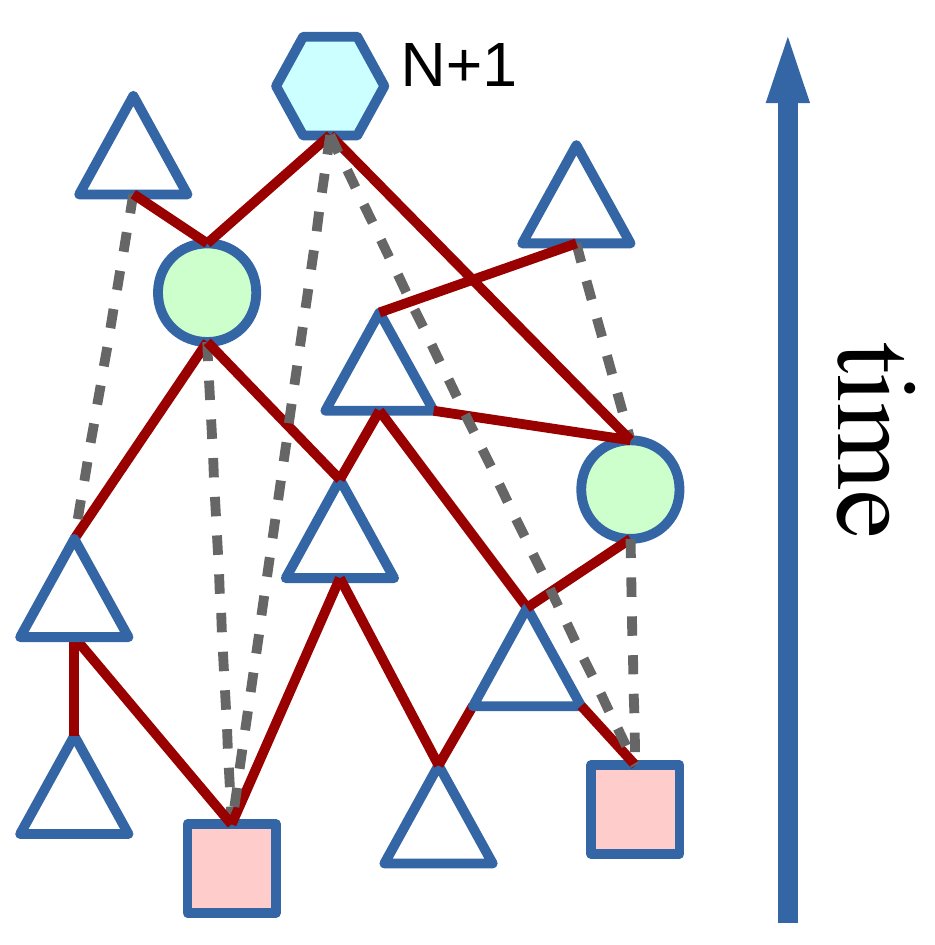}
\end{center}
\caption{An illustration of the transitive reduction to identify citations to `core' papers.  On the left is the citation network of figure \ref{Modelc} after the new paper $(N+1)$ (blue hexagon) and its references (green circles for its core references, red squares for its subsidiary references) have been added. On the right is the transitive reduction of this network where the dashed grey edges are the ones removed under transitive reduction. In this case, after transitive reduction the new paper $(N+1)$ is only linked to the core papers in its references.  The declustering method described in the text produces the same result in this case, but not in general.}
%Note if you want to reference in the caption \protect{\cite{citationlabel}}.
\label{Modelcnew}
\end{figure}

To find these references from papers to their core papers in the hep-th network we used a declustering method, a quicker and simpler version of transitive reduction \cite{clough2013transitive,CE14}. For each paper $X$ we start from the oldest paper referenced.  We delete any references from nodes $X$ to $Z$ if there exists a reference route from $X$ to $Y$ and $Y$ to $Z$, i.e.\ we remove the long edge of any triangles found between paper $X$ and its references, figure \ref{Modelcnew}. This will definitely remove all of the links from a paper to its subsidiary papers, but it is possible to remove other types of links. While not perfect, we use this to get an estimate for the average number of citations to core papers in the hep-th data, and we found this to be 3.9\footnote{Note that the average in-degree of the full un-reduced arXiv network is 12.82. The average in-degree after two-step declustering is much less, 3.9. The fully transitively reduced network has an even smaller average in-degree of 2.27 \cite{GoldC4, clough2013transitive}, as expected.}.  This is approximately what we expect as we expect that an author references a few core papers, on average, and a proportion of the core papers' bibliographies $q$. By running the same process on the output from model C, using parameters where $\AS=200$ papers and line of $(p,q)$ values in figure \ref{contour} from the line giving the correct $z$ value, we determine that $p=0.55$ and $q=0.2$ are the best values.

\begin{figure}[h] %[H]
\centering
\includegraphics[width=0.55\linewidth]{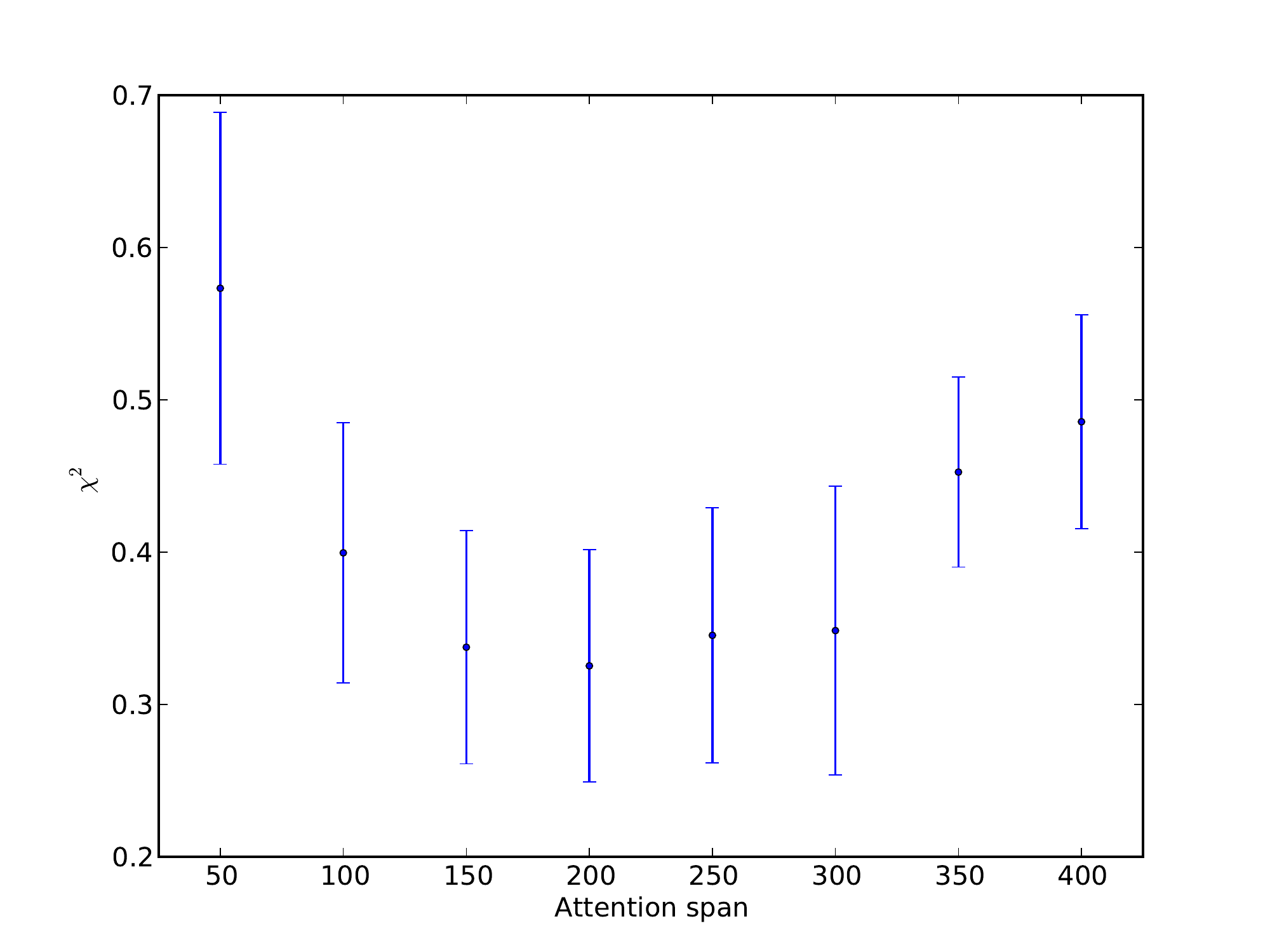}
\caption{Plot of $\chi^2$  (the squared difference between the $\sigma^2$ of the data and the model C for fixed $(p,q)=(0.55,0.20)$) against varying attention span $\AS$ measured in papers. Results show the mean and standard deviation for 10 runs of model C at each parameter value. The minimum is at at $200\pm 50$ papers.}
\label{minchi2}
\end{figure}

At this point we recall that our analysis of the fraction of zero cited papers $z$ which we used to choose the attention span $\AS$ is actually relatively insensitive to changes in $\AS$ when $q=0.2$.  So to find the optimal $\AS$ value for our best choice of $(p,q)=(0.55,0.2)$ we now look at the $\sigma^2$ values produced by model C. As before\footnote{For model C our final comparison tool is the $\sigma^2$ plot of model C and the hep-th data. Note that so far we have used $z$ and the number of core papers $C$ as our comparison tools.}, to compare model C to the data we use $\chi^2$ of equation \tref{eq:Bchi2} to make the comparison, as shown in figure \ref{minchi2}. From this we find a minima at $\AS=200$ papers, and so this is our optimal value for attention span $\AS$.

% ------------------------------------------------------------------------------------------
\subsection{Checks on the Optimal Values}

We have arrived at an optimal set of values for our model C: $p=0.55$, $q=0.20$ and $\AS=200$ papers. It is worth looking to see how these values compare with what we know from elsewhere.

For the processes described by model C we might expect (assuming no correlation between in- and out-degree)\footnote{This estimate assumes no correlation between in- and out-degree as a better estimate for the average in-degree of core papers chosen using cumulative advantage is $\texpect{(\kin)^2}/\texpect{\kin}$.} that
\begin{equation} \label{eqfindq}
\texpect{\kout}= C (1+q \texpect{\kout})
\end{equation}
where $C=3.9$ is the average number of core papers in a bibliography and $\texpect{\kout}=12.0$. This gives us $q=0.17$ which is very close to the value we extracted numerically.

Our value for $q$ also compares well with values used by Simkin and Roychowdhury \cite{SR05a} in their model and their different data set. They create a simple model in which a new publication references a few core papers and a quarter of the core papers' bibliographies. Therefore, they expect 25\% of a core paper's bibliography to be copied which is not too different from the 20\% we have arrived at for model C. Simkin and Roychowdhury also studied errors in bibliographies \cite{SR05b}. Using a statistical model for this process they again arrive at the conclusion 70-90\% of the references to papers are literally copied from existing papers.  We find this value to be, on average, $Cq\texpect{\kout}/ \texpect{\kout} = Cq \approx 0.67\approx 0.70=70\%$ or $0.78\approx 80\%$ from the second term on the right side of equation \ref{eqfindq}, for $R=3.9$ and $q=0.173$ or $q=0.20$ (which are the mathematically expected value of $q$  and the $q$  derived from fitting the model C to the hep-th data, respectively). Both values are compatible with our choice of $q=0.20$\footnote{We have also done another check. The proportion of core papers referenced per new publication is therefore, on average $=\frac{C}{\texpect{\kout}}=\frac{C}{C(1+q\texpect{\kout})}=\frac{3.9}{12.0} \approx 0.3$ for $q$=0.17 or $q$=0.20 (which are the mathematically expected value of $q$  and the $q$  derived from fitting the model C to the hep-th data, respectively). In \textit{A Mathematical Theory of Citing} and Price \cite{P65} this was found to be 0.1 and 0.15 for their models, respectively. Again, our values are consistent with these as they are low.}. Analysis of similar arXiv data sets using the different technique of transitive reduction by Clough et al.\ \cite{clough2013transitive} is also consistent with these values.

The value of the attention span model parameter $\AS=200$ is a fraction of a year (a year is 2000 papers in rank time) and therefore at first appears to be surprisingly low. This result may be necessary if a low $\AS$ leads to the larger $\sigma^2$ values. The idea is that if a paper has not gained citations in this short time span, then it is very unlikely to gain many citations later. On the other hand, if a paper gains a few citations in this short time, then they are likely to be referenced again and again as a citation to a subsidiary paper in the future publications.  Not only does this gives us the `rich-get-richer' effect and so the long-tail of the in-degree distribution, more importantly the short attention span is exacerbating the difference between high and low citations, exactly the type of effect we need to widen the distribution of citations for papers published in the same year, i.e.\ it makes $\AS$ higher. Note that within the literature an attention span of less than a year is also found \cite{AMTC}.

Another issue could be that the academic field covered by the hep-th arXiv, largely the string theory approach to particle physics, has an unusually short time scale.  With no experimental apparatus, theoretical physics can react quickly --- a paper can be written in as little as  week from an initial idea.  Out impression is also that approximate results or swift reactions to the latest scraps of experimental information are acceptable in hep-th.  So again the culture is one of speed rather than low publishing rate and more contemplative approach often associated with pure maths, a subject with strong ties to string theory.  To see our low attention span value reflects particular properties of the hep-th area, we would need to extend our approach to different fields of research. This attention span is

%\newpage

\begin{figure}[h] %[H]
\centering
\includegraphics[trim = 0cm 0.5cm 0cm 1.5cm, clip = true, scale=0.45]{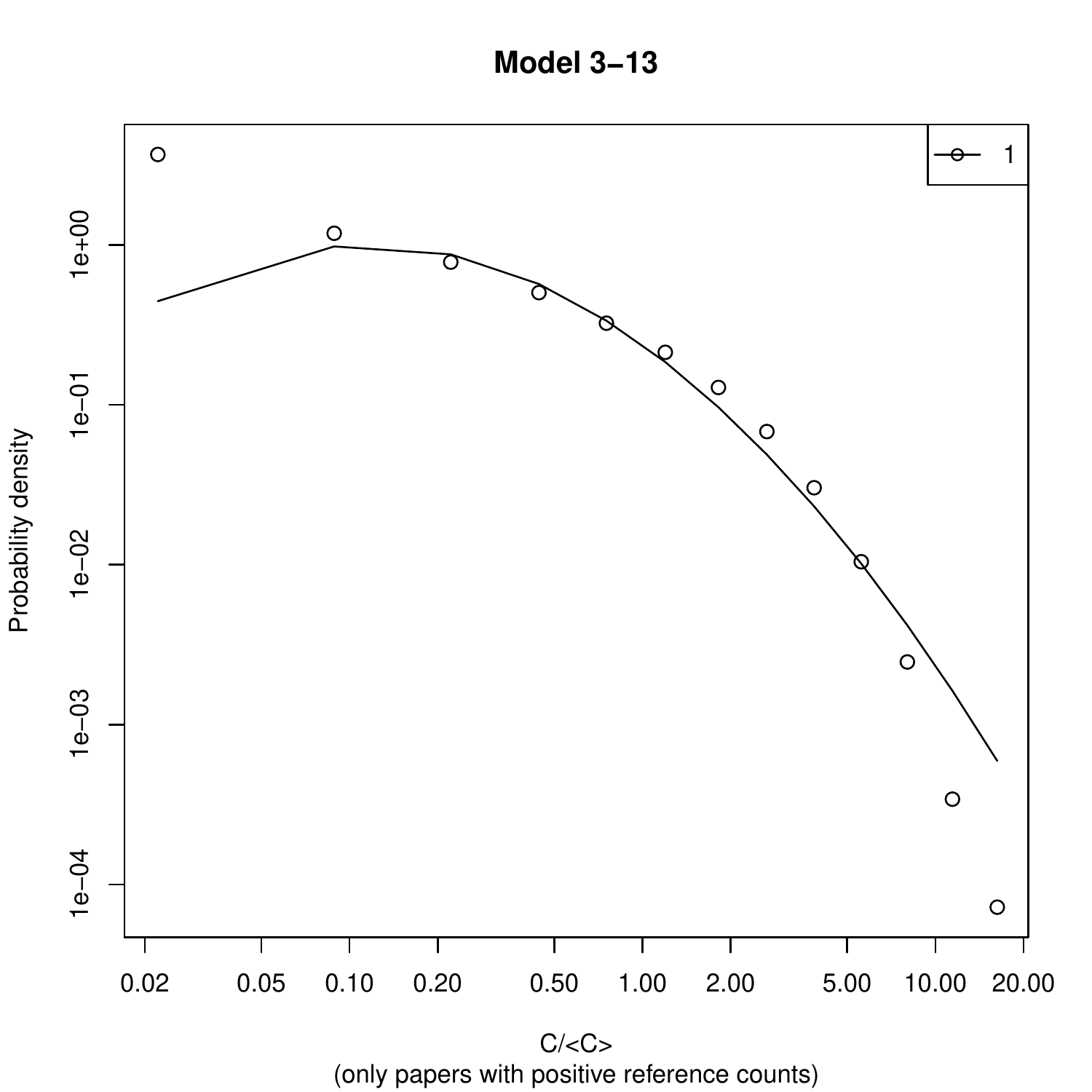}
\caption{This is the citation distribution for all papers for final model C, $(p,q)$ =$(0.55,0.20)$ and $\AS$ =200 papers, plotted against normalised citation count, on a log-log plot. The tail has a large with of $1.68\pm 0.27$ consistent with the data $1.78 \pm 0.14$. The fitted lognormal reaches a normalised citation count of approximately 50, the same as the hep-th data, figure \ref{dataindeg}, and 2/3 times that of model A and B, see figures \ref{indegma} and \ref{indegMB} respectively.}
\label{SHindeg}
\end{figure}

\begin{figure}[h] %[H]
\centering
\includegraphics[scale=0.38]{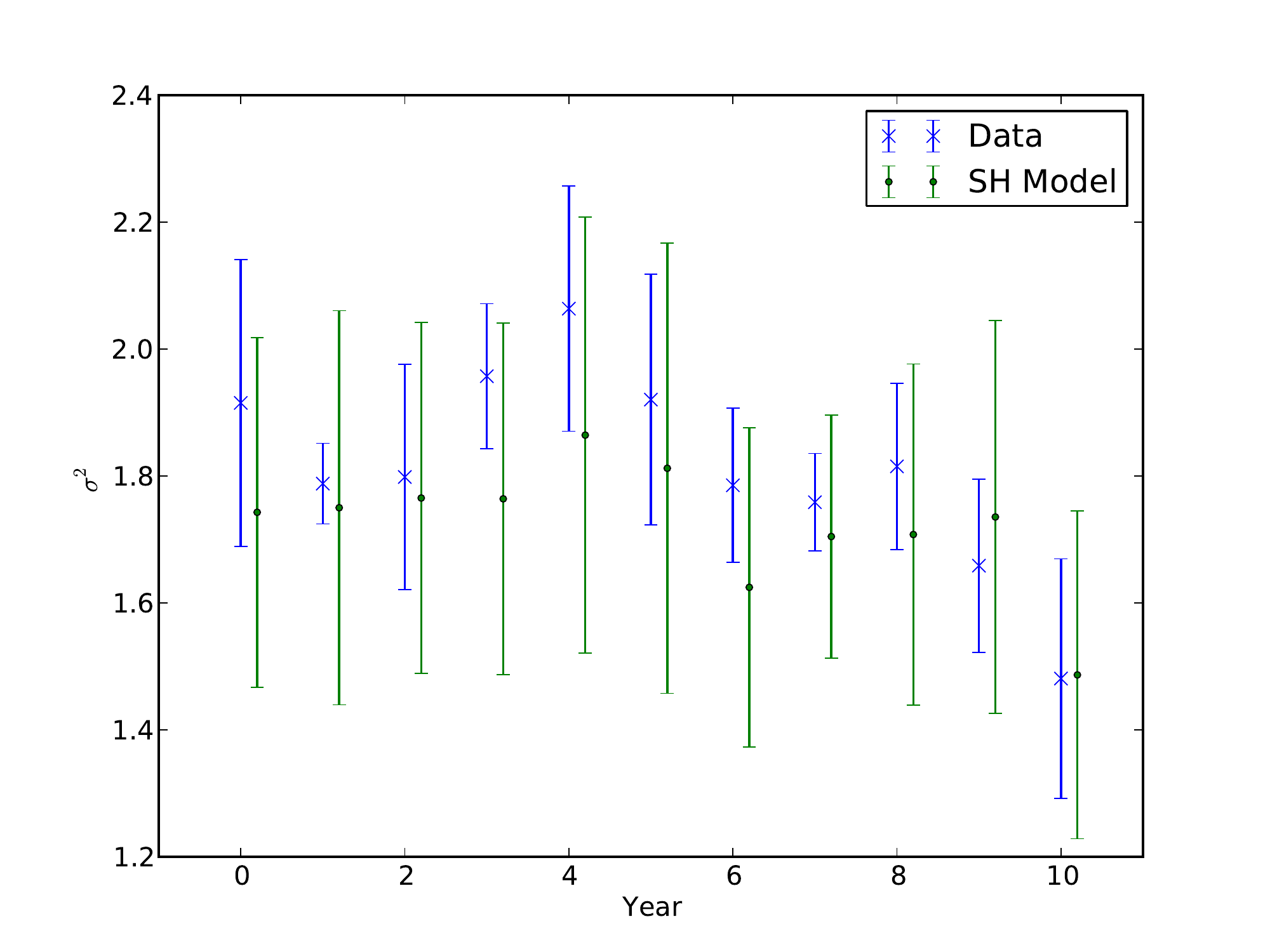}
\includegraphics[scale=0.38]{{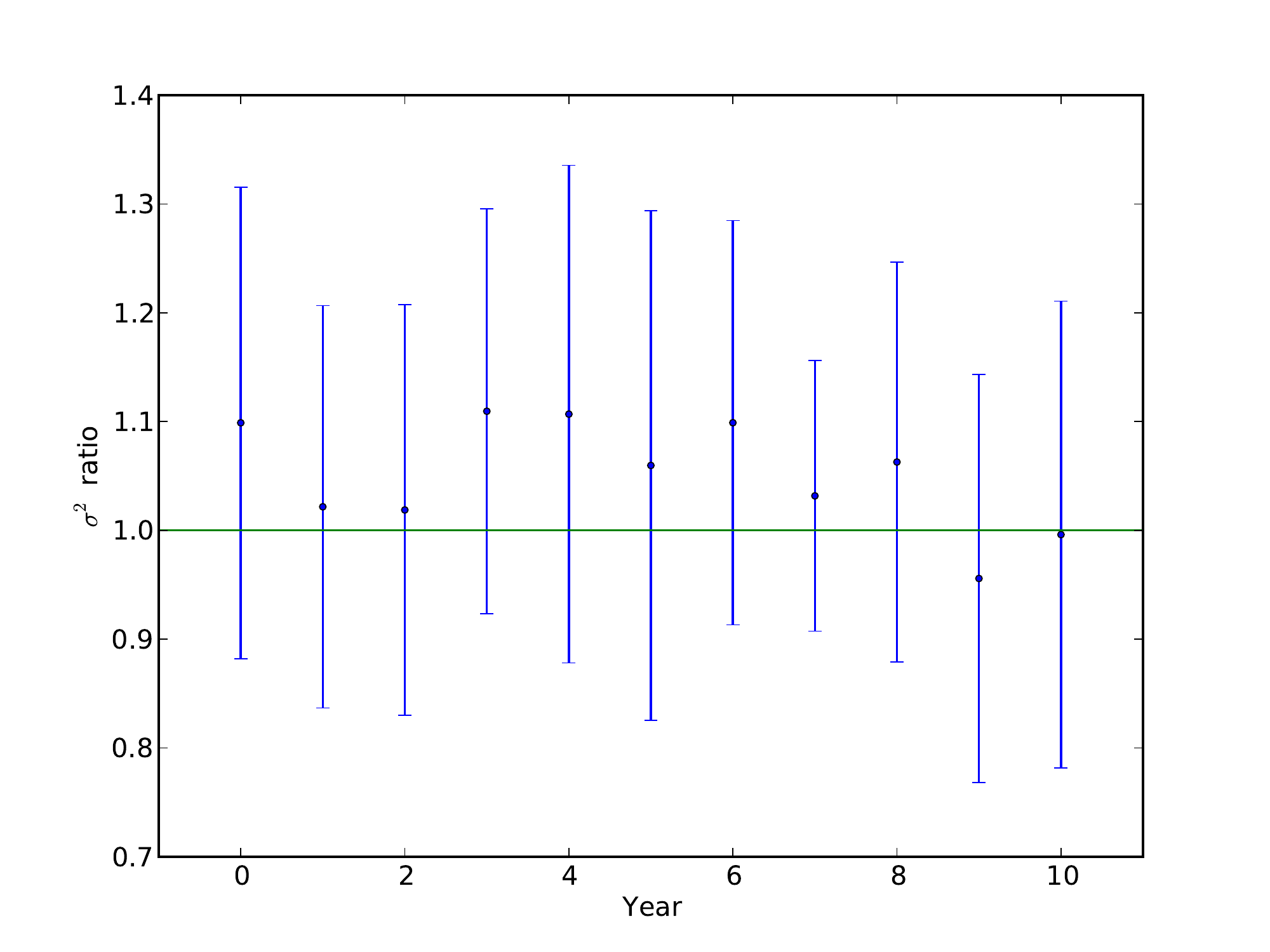}}
\caption{Plots of width of the long-tail citation distribution, $\sigma^2_y$, for each year for data and model C. On the left
are the absolute values for the $\sigma^2_y$ values for model C (green) and for the hep-th data (blue). On the right is a plot of the ratio $\sigma^2_\mathrm{y,data}/\sigma^2_\mathrm{y,model}$, the $\sigma^2$ value from the hep-th data divided by that found in model C with optimal parameters. The data and model C results for $\sigma$ are consistent with each other, being within $\pm 0.2$ of one another and within each others error bars. The data's $\sigma^2$ values are slightly higher for years 0 to 8 (1992 - 2000) but slightly lower for years 9 and 10 (2001 and 2002) though differences are not statistically significant.
}
\label{SHstoratio}
\label{SHsto}
\end{figure}

We can now look at the output from our model C with the optimal parameter values, $(p,q)=(0.55,0.20)$ and $\AS$ =200 papers.  For the citation  distribution for the whole time period, shown in figure \ref{SHindeg}, we find a long-tailed distribution with $\sigma^2 = 1.68\pm 0.27$, consistent with the hep-th data $1.78 \pm 0.14$. Looking at the shape of the citation distribution for publications in one year, the widths, $\sigma^2_y$ are all very consistent with the data, see figure \ref{SHsto}. That is we find the large and constant widths $\sigma^2_y \approx 1.8$ seen in the data but which were not seen in model A (see figures \ref{indegma}) nor in model B (see figure \ref{indegMB}).

% ******************************************************************************
\section{Conclusions} \label{conc}

We have created three citation network models and compared them to the hep-th arXiv data. Our aim has been to find what processes are sufficient to produce long-tailed citation distributions for papers published in one year which are of similar width when rescaled by the average citation count in that year.  That is for each year there are always a large number of papers cited only a fraction of the average for that year, but there are also a few but significant number of papers in the same year which have a much larger number of citations relative to the average.  We quantify this width via a $\sigma^2$ parameter defined using a lognormal fit to the tail of the distribution.

The difficulty in achieving our aim is illustrated by our first two models.  Model A was a simple variation of the Price model, with references added to new papers with a mixture of cumulative and uniform random attachment. Model B added a time decay factor so it was more likely to reference a paper over a recent time scale, $\thalf$. In both cases a long tailed citation distribution was produced for papers published in one year, but the range of citation counts, the width of this tail, was far too small. On the other hand our model C with just three parameters used a mixture of global information with a time decay factor, along with local searches, to produce distributions with the right type of citation distribution shapes. Thus one of our major conclusions is that both local searches \emph{and} global information have a noticeable influence on citation distributions. In particular our model and data suggest that around 70\% to 80\% of papers cited are `subsidiary papers', that is papers which are also cited by the other papers in the bibliography, the `core papers'.  Similar results have been seen by Simkin and Roychowdhury \cite{SR05b} who use a similar model but different data and methods to arrive at this result. From a different perspective, Clough et al.\ \cite{clough2013transitive} have also shown that around 80\% of citations are removed from arXiv citation networks through the process of transitive reduction (an extended version of the declustering algorithm used here).  The ``essential links'' left in a citation network after transitive reduction \cite{CE14} are, approximately, just the citations to core papers in model C. Clough et al.\ \cite{clough2013transitive} also show that these remaining essential links to core papers tend to be over short time scales.

Our approach has a number of key features.  We study the citation distribution of each year, representing it in terms of the fraction of zero-cited papers and the width of the fat-tail of the citation distribution\footnote{Note that the average citation count of model C is also compatible with the hep-th data}. We also looked at the average time needed to acquire half of the citations for papers published in one year in order to choose the time scale in the model.  Finally we found that working in rank time rather than calendar year was of great use in model making. This combination has not been used before.  The time scales provided some particularly interesting results.

One aspect our work suggests is worth considering in the future is the role of the length of bibliographies, the out-degree distribution.  We used an unrealistic normal distribution here though at least this gives fluctuations in $\kout$, something not present in some other models \cite{AMTC, ISI:000187183600031}. Our preliminary investigations found the shape of the distribution of bibliography length did change the citation distribution. For instance, when using a lognormal distribution as the out-degree distribution instead of the normal distribution in the model C, we found that the $\sigma^2$ plots did change, though results were broadly similar. It would be interesting to find a simple model with realistic distributions for both in- and out-degree distributions and what factors determine the out-degree for new publications.

%****************************************************************************************
\section*{Acknowledgements}

We would like to thank James Gollings and James Clough for allowing us to use their transitive reduction code from which we created our own declustering code. We would like to thank Tamar Loach for sharing her results on related projects and M.V.~Simkin for discussions about his work.

%\bibliographystyle{ieeetr}
%%\bibliographystyle{abbrv}
%%\bibliography{RefsFinal}
%\bibliography{citationmodel}

\newcommand{\ttitle}[1]{#1}
\newcommand{\tjournal}[1]{\textit{#1}}
\newcommand{\tvol}[1]{\textit{#1}}

% AAAAAAAAAAAAAAAAAAAAAAAAAAAAAAAAAAAAAAAAAAAAAAAAAAAAAAAAAAAAAAAAAAAAAAAAAAAAAAAA
\appendix

\section{Appendix}

\subsection{Fitting Procedure}

We follow the procedure used in \cite{EHK12,Goldberg}. We use logarithmic binning so that the citations in bin b $c_b \in \mathbb{Z}$ with $c_{b+1}$ equal to $R c_b$ rounded to the nearest integer or to $(c_{b}+1)$, whichever is the highest, where $R$ is some fixed bin scale chosen to ensure there are no empty bins below the bin containing the highest citation values. The edge of the first bin is chosen to be the lowest integer above the value $0.1 \texpect{c}$. In order to make the fit we compare the total value in the $b$-th bin,  $n_b= \sum_{c=c_{b}}^{c_{b+1}} n(c)$, against the expected value
\begin{eqnarray}
 n_b^{(\mathrm{expect})}
 &=&
 (1+A) N \int_{c_{b}-0.5}^{c_{b+1}+0.5}  dc \; \frac{1}{\sqrt{2\pi} \sigma c }
 \exp\left\{ -\frac{(\ln(c/\texpect{c})+(\sigma^2/2)-B)^2}{2\sigma^2 } \right\}
 \, .
 \label{lognormalbin}
\end{eqnarray}
This gives us a sequence of data and model values which are compared using a non-linear least squares algorithm to give us values for $\sigma^2$, $A$ and $B$.

% --------------------------------------------------------------------------------
\subsection{Lognormal out-degree Distribution}\label{logout}

\begin{figure}[h] %[H]
\centering
\includegraphics[width=0.55\linewidth]{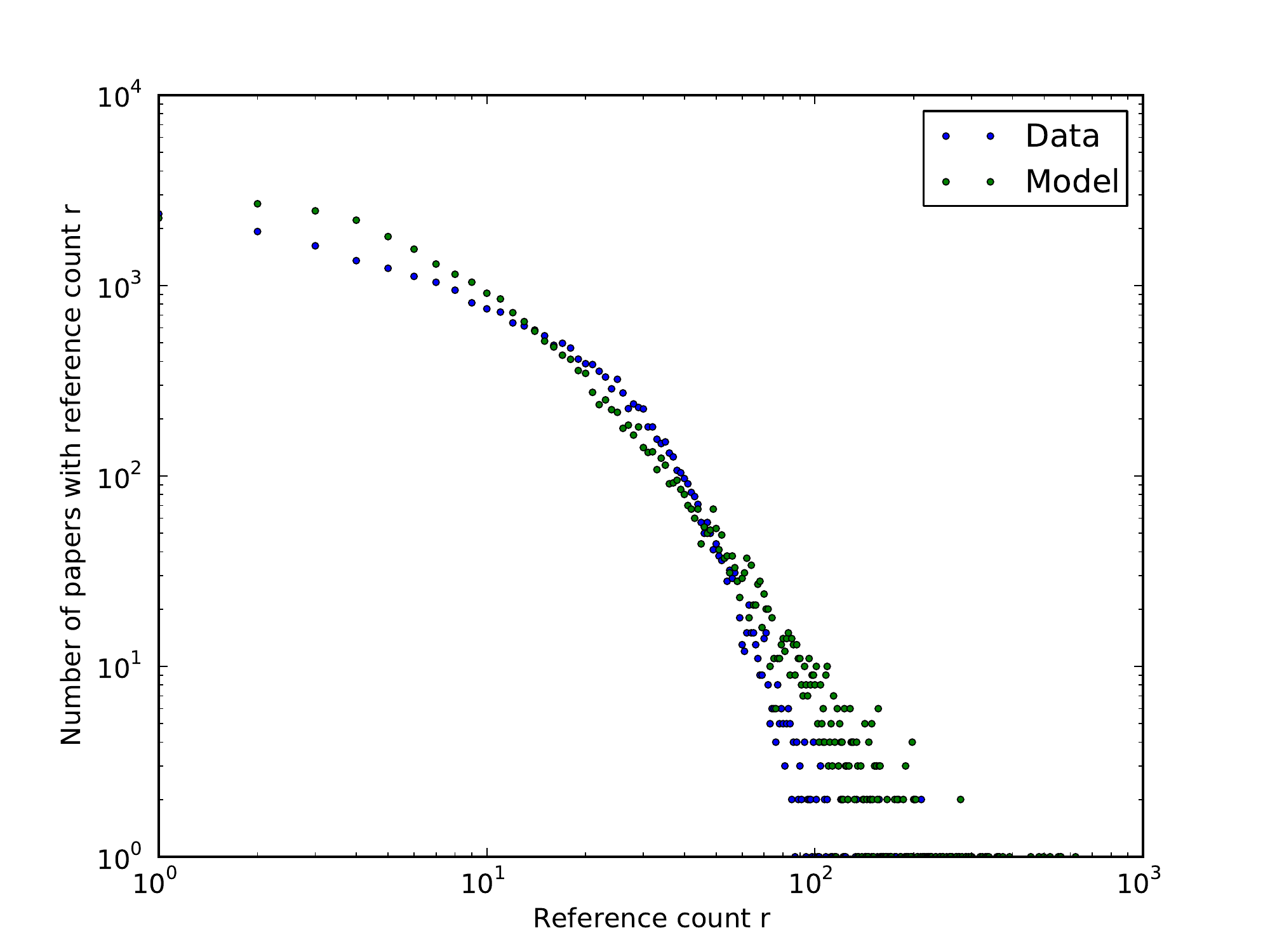}
\caption{This is a plot of the out-degree distribution of 27,000 publications from the hep-th arXiv data, in blue, on a log-log plot. Superimposed, in green, is a plot of 27,000 numbers generated by the lognormal distribution fitted to the  out-degree of the hep-th arXiv data. We observe that the lognormal is a better fit to the data than the normal in figure \ref{fig:model_and_data_outdeg}.}
\label{fig:model_and_data_outdeg_lognorm}
\end{figure}

In all above models the citation networks were created by determining the number of references a new node would create (via a normal distribution mean 12.0, standard deviation 3.0 references) and then having a method of deciding which nodes to reference. However, we found that a lognormal fits the out-degree distribution of the hep-th data better than a normal distribution, figure \ref{fig:model_and_data_outdeg} and \ref{fig:model_and_data_outdeg_lognorm}, respectively.

\begin{figure}[h] %[H]
\centering
\includegraphics[width=0.55\linewidth]{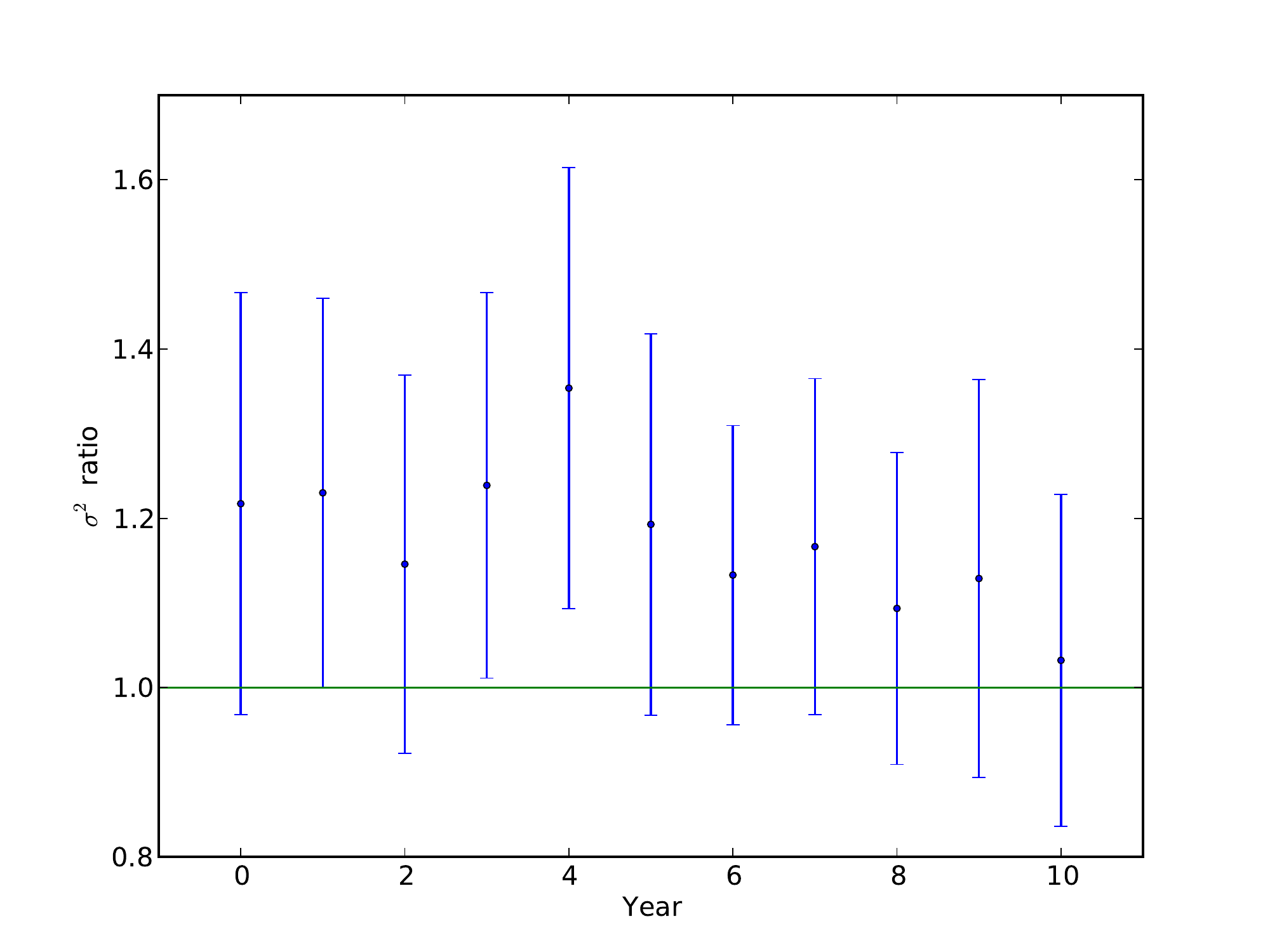}
\caption{This is the ratio of $\sigma^2$ associated with the in-degree distribution of papers published in the same year for the data (years from 1992, 1993 etc. relabelled to 0, 1 etc.) divided by that of the modified model C (where the out-degree is determined by a lognormal distribution, above). The plot and error bars lie within 1.0 therefore the modified model C is consistent with the data. Therefore the modified model C is promising, a significant improvement on model A and B, figures \ref{sigma2dvma} and \ref{Bsigma2}, respectively. However, the data is always lower than the modified model C; the points are always above the 1.0 line and not as close to 1.0 as the model C which implies the need for modification of the parameters in model C. So modifying the out-degree does change the in-degree, which contradicts \cite{citeulike:1007417}. We conjecture that by changing the attention span parameter this model's $\sigma^2$ plot could increase to match the data.}
\label{fig:sigma2_modelC_lognorm}
\end{figure}

As further work we inputted this fitted lognormal to determine the number of references created by a new node into the model C and ran it for our final parameters $(p,q)$$=(0.55,0.20)$ and $\AS$ =200 papers. The ratio of the $\sigma^2$ values associated with the in-degree distribution of papers published in the same year for the data is divided by the corresponding year's $\sigma^2$ for this modified model C and plotted against year in figure \ref{fig:sigma2_modelC_lognorm}. We find that the ratio is close to 1, however, it is not as close as the original model C, figure \ref{SHsto}. Therefore the $\sigma^2$ plot \textit{does} depend on in-degree, contradicting \cite{ISI:000302837300019}, who say it is `innocuous' to the in-degree distribution of the citation network. Although this out-degree distribution has been observed by the literature \cite{citeulike:1007417} its use in a citation network model is novel and original.

Although the $\sigma^2$ plot is lower than that of the original model C we conjecture that by varying the $\AS$  of the model C the $\sigma^2$ plot could match that of the data's, this may also increase the attention span to something closer to a year as expected by \cite{AMTC}.

\end{document}